\documentclass[%
reprint,
bibnotes,
 amsmath,amssymb,
aip,
longbibliography
]{revtex4-2}

\usepackage{upgreek}
\usepackage{graphicx}
\usepackage{dcolumn}
\usepackage{bm}
\usepackage{xcolor}
\usepackage[pdftex, colorlinks=true, urlcolor=blue, linkcolor=blue, citecolor=blue, pdfstartview=FitH, plainpages=false]{hyperref}
\usepackage[utf8]{inputenc}

\newcommand{\kk}{\mathbf{k}}
\newcommand{\qq}{\mathbf{q}}
\newcommand{\qnu}{{{\qq\nu}}}
\newcommand{\wqnu}{\omega_{\qnu}}
\newcommand{\Dqnu}{D_{\qnu}}
\newcommand{\LL}{\mathrm{L}}
\newcommand{\KK}{\mathrm{K}}

\begin{document}

\preprint{AIP/123-QED}

\title{Indirect light absorption model for highly strained silicon infrared sensors}

\author{Nicolas Roisin}
\email{nicolas.roisin@uclouvain.be}
\affiliation{ Institute of Information and Communication Technologies, Electronics and Applied Mathematics (ICTEAM), Université Catholique de Louvain, Place du Levant,3 L5.03.02, B-1348 Louvain-la-Neuve, Belgium}
\author{Guillaume Brunin}
\affiliation{ Institute of Condensed Matter and Nanosciences (IMCN), Université Catholique de Louvain, Place Louis Pasteur, 1 bte L4.01.06, B-1348 Louvain-la-Neuve, Belgium}
\author{Gian-Marco Rignanese}
\affiliation{ Institute of Condensed Matter and Nanosciences (IMCN), Université Catholique de Louvain, Place Louis Pasteur, 1 bte L4.01.06, B-1348 Louvain-la-Neuve, Belgium}

\author{Denis Flandre}%
\affiliation{ Institute of Information and Communication Technologies, Electronics and Applied Mathematics (ICTEAM), Université Catholique de Louvain, Place du Levant,3 L5.03.02, B-1348 Louvain-la-Neuve, Belgium}
\author{Jean-Pierre Raskin}%
\affiliation{ Institute of Information and Communication Technologies, Electronics and Applied Mathematics (ICTEAM), Université Catholique de Louvain, Place du Levant,3 L5.03.02, B-1348 Louvain-la-Neuve, Belgium}

\date{\today}

\begin{abstract}

The optical properties of silicon can be greatly tuned by applying strain, opening new perspectives particularly in applications where infrared is key. 
In this work, we use a recent model for the indirect light absorption of silicon and include the effects of tensile and compressive uniaxial strains. 
The model is based on material properties such as the band gap, the conduction and valence bands density-of-states effective masses, and the phonon frequencies, that are obtained from first principles including strain up to $\pm2$\% along the [110] and [111] directions. 
We show that the limit of absorption can increase from 1.14~$\upmu$m (1.09~eV) to 1.35~$\upmu$m (0.92~eV) under 2\% strain, and that the absorption increases by a factor of 55 for the zero-strain cutoff wavelength of 1.14~$\upmu$m when a 2\% compressive strain is applied in the [110] direction. 
We demonstrate that this effect is mainly due to the impact of strain on the electronic band gaps of silicon, directly followed by the valence band density-of-states effective mass. 

\end{abstract}

\maketitle

\section{Introduction}
\label{sec:intro}

Silicon is a well-known material in the electronic industry. 
It has been extensively used over the past decades to develop ever more efficient electronic devices. Compared to other materials used in infrared applications, e.g., germanium, indium or gallium,\cite{Licht2015} silicon devices present the advantages of low toxicity and environmental impact with straightforward integration for CMOS circuits.
Most improvements were realized by reducing the device dimensions.\cite{Warnock2011} 
However, reducing dimensions has physical limits, and other technical innovations have been needed. \cite{Waldrop2016} 
Strained silicon has been suggested and indeed quickly showed promising properties such as improved electron or hole mobilities.\cite{Jaeger2018,Heremans2016} 
This technique is now widely used in industry especially to enhance Si-based CMOS performances with a mobility increase up to 90\% with stressors or epitaxial growth.\cite{Reiche2009}
Today, strain engineering is used routinely to manufacture semiconductor devices, in order to boost performances at low additional cost.\cite{Gnanachchelvi2016,Thompson2006,Kleimann1998} 
Even though the effects of small strain on the electronic properties of silicon have already been studied, both theoretically and experimentally,\cite{Nielsen1985,Rieger1993,Goroff1963} highly strained silicon remains largely unexplored.
Only recently, it was shown how to reach high strain in different structures based on nanoribbons\cite{Passi2012,Bhaskar2012,KumarBhaskar2013}, nanowires\cite{Zhang2016} or nanomembranes.\cite{Cavallo2012,Montmeat2016}
The interest for highly strained silicon is therefore enhanced since such strain levels can significantly change its optical properties and make silicon a suitable material for infrared applications.\cite{Munguia2008,Lange2016,Schriever2012,Cazzanelli2012,Fischetti1996,Wen2015}

The indirect band gap of silicon enables light absorption through interband transitions up to the near-infrared region, for photon energies roughly larger than the fundamental band gap of 1.12~eV at 300\,K (wavelengths lower than 1.1~$\upmu$m).\cite{Lin2017} 
Such indirect transitions require additional momenta provided by the lattice vibrations, the so-called phonons.\cite{Noffsinger2012} 
This phonon-assisted process dominates the absorption spectrum of silicon in the energy region between the fundamental and the direct (3.2~eV) band gaps. 
For energies larger than the direct band gap, the absorption quickly increases and is mainly dictated by direct interband transitions. 
Despite the importance of the subject, only a few theoretical computations of the indirect absorption spectrum of silicon exist,\cite{Noffsinger2012,Zacharias2015,Tsai2018,Patrick2014} 
mostly because of the complexity of the interactions between electrons and phonons.\cite{Giustino2017} 
Only recently, first-principles computations of the phonon-assisted absorption were able to quantitatively describe the absorption spectrum of silicon.\cite{Noffsinger2012,Zacharias2015} 
However, while these first-principles computations give accurate results, the computational cost is very large.

In this work, we extend the theoretical model of Tsai\cite{Tsai2018} developed for relaxed silicon to the highly strained crystal case. 
The model requires only a few parameters such as the direct and indirect band gaps, the conduction and valence bands density-of-states effective masses (i.e., the effective masses leading to the same density of states as if the bands were parabolic), the phonon frequencies, and the deformation potentials for the different electronic transitions. 
Using first-principles computations, we show that the main impact of strain is on the band gaps and the effective masses, while the phonon energies remain almost unchanged.
The advantage of this model compared with fully first-principles computations is the time required to obtain the full spectrum, particularly when many strain levels have to be considered. 
Using our model, we demonstrate that the cutoff wavelength of silicon, defined here as the wavelength above which the absorption becomes lower than $1$~cm$^{-1}$, can increase from 1.14 to 1.35~$\upmu$m (decrease from 1.09 to 0.92~eV) under a 2\% tensile or compressive strain.
Additionally, we show that the absorption coefficient increases by a factor of 15 to 55 at the zero-strain cutoff wavelength, therefore largely increasing the efficiency of silicon-based devices in the infrared.

\section{Indirect light absorption model}
\label{sec:model}

Light absorption can be due to direct and indirect processes. 
In this work, we focus on the indirect light absorption, which is the dominating process for photons of energies below 3.2~eV (above 387 nm) in relaxed silicon. 
In this case, the absorption coefficient $\alpha$ is usually expected to vary as the square of the incoming photon energy $E_\lambda$, with $\lambda$ the photon wavelength. More precisely, assuming single-phonon processes, the absorption coefficient can be expressed as\cite{Pankove1975}
\begin{equation}
    \begin{split}
    \alpha(E_\lambda) = & P \displaystyle\left[ 
            \frac{\left(E_{\lambda} - E_g + E_{ph}\right)^2}{\exp\left(\frac{E_{ph}}{k_BT}\right) - 1} H(E_{\lambda} - E_g + E_{ph}) \right. \\
            & \displaystyle\left. + \frac{\left(E_{\lambda} - E_g - E_{ph}\right)^2}{1 - \exp\left(-\frac{E_{ph}}{k_BT}\right)} H(E_{\lambda} - E_g - E_{ph})\right],
    \end{split}
\label{eq:alphaind}
\end{equation}
where $P$ is a material-dependent coefficient, $E_g$ is the fundamental band gap, $E_{ph}$ is the phonon energy, $k_B$ is the Boltzmann constant, $T$ is the temperature and $H(x)$ is the Heaviside unit-step function. The two terms in Eq.~\eqref{eq:alphaind} correspond to  phonon absorption and emission processes, respectively. 
This equation can correctly describe the general trend of the absorption spectrum.\cite{RAJKANAN1979} It can also be used to extract band gap values (the absorption coefficient being very low for photon energies below the fundamental band gap).\cite{Pankove1975} In this relation, $P$ is treated as a fitting parameter of the semi-empirical model. Therefore, this \textit{a priori} unknown proportionality coefficient prevents a fully theoretical prediction of materials performances, particularly in a case such as highly strained silicon for which there are no experimental data available. 

\begin{figure}
    \centering
    \includegraphics[width=1\linewidth]{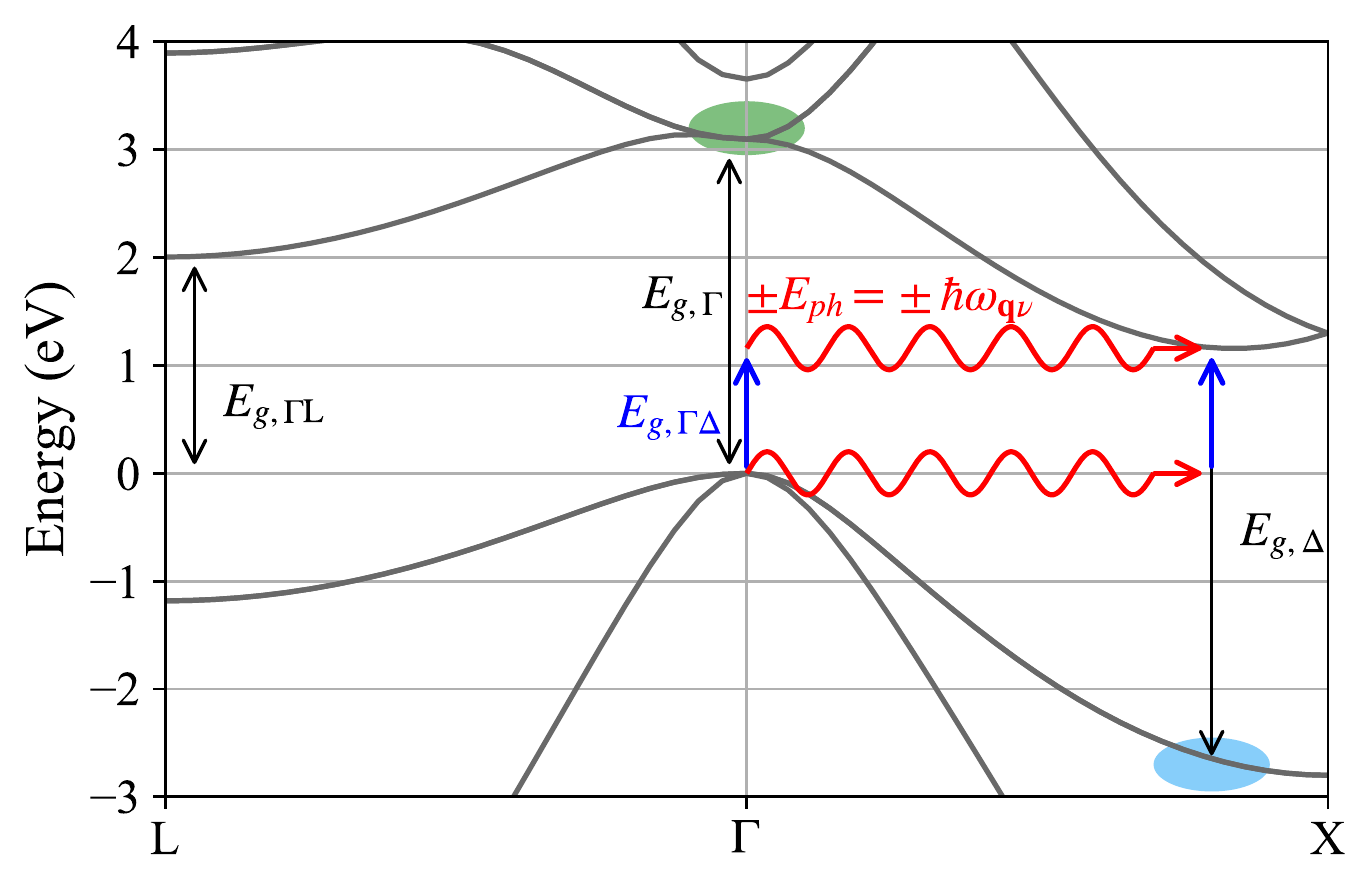}
    \caption{Electronic band structure of silicon. Two possible absorption routes for the indirect light absorption are represented by blue (photon absorption) and red (phonon absorption or emission) arrows. The phonon energy is $E_{ph} = \hbar\wqnu$. The intermediate states are highlighted by green and blue areas, and the important band gaps $E_g$ for indirect transitions between different $\kk$ points are represented by vertical arrows. The zero of energy is located at the valence band maximum.}
\label{fig:bandtransition}
\end{figure}

Recently, Tsai derived a theoretical model for computing the indirect absorption coefficient based on second-order time-dependent perturbation theory.\cite{Tsai2018} 
The model extends Eq.~\eqref{eq:alphaind} by considering all the possible types of intervalley scattering processes for the indirect transitions. 
Some of these transitions are represented in Fig.~\ref{fig:bandtransition} with the band structure of relaxed silicon. 
Such indirect light absorption happens by either the absorption of a photon (represented by blue arrows) followed by the absorption or emission of a phonon (represented by red wavy arrows), or vice versa. 
For each valley of the conduction band where the electrons might be scattered, 
two intermediate states are considered. 
The first one (in green in Fig.~\ref{fig:bandtransition}) is located at $\Gamma$ in the conduction band, and the second one (in blue in Fig.~\ref{fig:bandtransition}) is located in the valence band, at the high-symmetry $\kk$ point characterizing the conduction band valley (denoted by $\Delta$ in the following), $\kk$ being the electron wavevector. 
For a given transition connecting two valleys through a phonon of wavevector $\qq$ and mode $\nu$, 
the absorption coefficient derived by Tsai is given by\cite{Tsai2018}
\begin{widetext}
\begin{equation}
\begin{split}
    \displaystyle \alpha_\qnu = & \frac{A N_\mathrm{val}}{E_\lambda \hbar\wqnu} \left[ \frac{N_\qq (E_\lambda - E_g + \hbar\wqnu)^2}{|E_{g1} - E_\lambda - i\Gamma_d|^2} H(E_\lambda - E_g + \hbar\wqnu) \right.
    + \frac{(N_\qq + 1) (E_\lambda - E_g - \hbar\wqnu)^2}{|E_{g1} - E_\lambda - i\Gamma_d|^2} H(E_\lambda - E_g - \hbar\wqnu) \\
    & \hspace{10.5mm} \displaystyle +\frac{N_\qq (E_\lambda - E_g + \hbar\wqnu)^2}{|E_{g2} - E_\lambda - i\Gamma_d|^2} H(E_\lambda - E_g + \hbar\wqnu)
    \left. +\frac{(N_\qq + 1) (E_\lambda - E_g - \hbar\wqnu)^2}{|E_{g2} - E_\lambda - i\Gamma_d|^2} H(E_\lambda - E_g - \hbar\wqnu)\right],
\end{split}
\label{eq:Tsai}
\end{equation}
\end{widetext}
where $N_\mathrm{val}$ is the number of degenerated conduction valleys, $E_g$ is the indirect band gap (fundamental or not, depending on the considered valley), $E_{g1}$ and $E_{g2}$ are the direct band gap energies providing intermediate states (for instance, $E_{g,\Gamma}$ and $E_{g,\Delta}$ in Fig.~\ref{fig:bandtransition}), $\Gamma_d$ is a damping factor, and
\begin{equation}
    N_\qq = \left(\exp\left(\frac{\hbar\wqnu}{k_BT}\right)-1\right)^{-1}
\end{equation}
is the Bose-Einstein distribution for the number of phonons of frequency $\wqnu$.
The material-dependent proportionality constant $A$ is defined as
\begin{equation}
    A = \frac{e^2 m_c^{3/2} m_v^{3/2} E_p \Dqnu^2}{96\pi^2 \hbar^3 m_0 c \varepsilon_0 n_r \rho_L},
\label{eq:Aind}
\end{equation}
where $e$ is the electron charge, $m_0$ is the free electron mass, $c$ is the speed of light in vacuum, $\rho_L$ is the material density, $n_r$ is the refractive index, $E_p$ is an energy parameter for the electron-photon interactions, $\Dqnu^2$ is the phonon deformation potential, and $m_c$ and $m_v$ denote the density-of-states effective masses of the conduction and valence bands, respectively. To the contrary of the semi-empirical model, Eq.~\eqref{eq:Tsai} is based on a fully detailed physical understanding of the absorption mechanisms that is essential to correctly predict the behavior of the material.

As it is observed in Fig.~\ref{fig:bandtransition}, different indirect transitions have to be considered in silicon, corresponding to different valleys of the conduction band. 
Each of them gives a different contribution to the total absorption coefficient,
\begin{equation}
    \alpha = \sum_{\qnu} \alpha_\qnu.
\label{eq:alpha_tot}
\end{equation}

\section{Methods}
\label{sec:method}

\subsection{Strain}

We consider two crystallographic directions for the uniaxial strain, namely [110] and [111]. 
These modify the unit cell of silicon according to the related strain tensors,
\begin{equation}
    \textnormal{\textbf{E}}_{[110]} = \frac{1}{2} \left( 
    \begin{array}{ccc}
        (\varepsilon_{\perp}+\varepsilon_{\parallel}) & (\varepsilon_{\perp}-\varepsilon_{\parallel}) & 0 \\
        (\varepsilon_{\perp}-\varepsilon_{\parallel}) & (\varepsilon_{\perp}+\varepsilon_{\parallel}) & 0 \\
        0 & 0 & 2\varepsilon_{\perp}
    \end{array}\right)
\end{equation}  
\begin{equation}
    \textnormal{\textbf{E}}_{[111]} = \frac{1}{3} \left(
    \begin{array}{ccc}
        (\varepsilon_{\perp}+2\varepsilon_{\parallel}) & (\varepsilon_{\perp}-\varepsilon_{\parallel}) & (\varepsilon_{\perp}-\varepsilon_{\parallel}) \\
        (\varepsilon_{\perp}-\varepsilon_{\parallel}) & (\varepsilon_{\perp}+2\varepsilon_{\parallel}) & (\varepsilon_{\perp}-\varepsilon_{\parallel}) \\
        (\varepsilon_{\perp}-\varepsilon_{\parallel}) & (\varepsilon_{\perp}-\varepsilon_{\parallel}) & (\varepsilon_{\perp}+2\varepsilon_{\parallel}) \\
    \end{array}\right),
\end{equation}  
where $\varepsilon_{\perp}$ is the uniaxial strain applied on the material in the [110] or [111] direction while $\varepsilon_{\parallel}$ is the induced perpendicular strain. 
For elastic deformations, $\varepsilon_{\parallel}$ is proportional to $\varepsilon_{\perp}$:
\begin{equation}
    \varepsilon_{\perp} = -D_{[ijk]} \varepsilon_{\parallel},
\end{equation}
with
\begin{equation}
    D_{[110]} = \frac{2C_{11} C_{44} + (C_{11} + 2C_{12}) (C_{11} - C_{12})}{4C_{12} C_{44}}
\end{equation}
and
\begin{equation}
    D_{[111]} = \frac{C_{11} + 2C_{12} + 2C_{44}}{C_{11} + 2C_{12} - 2C_{44}},
\end{equation}
where $C_{11}=165.77$~GPa, $C_{12}=63.93$~GPa, and $C_{44}=79.62$~GPa are the elastic constants of silicon.\cite{Madelung1991}

\subsection{Computational methods}

For relaxed and strained silicon, different parameters are computed from first principles using density functional theory (DFT), as implemented in ABINIT.\cite{Gonze2020,Romero2020} 
We use norm-conserving pseudopotentials from the \textsc{PseudoDojo},\cite{Vansetten2018} in the generalized-gradient approximation (GGA) from Perdew-Burke-Ernzerhof (PBE) with a plane-wave cutoff of 20 Ha. 
For each strain level, we first perform a structural relaxation for the two atoms of the strained (fixed) unit cell. 
The electronic band structure is then determined using an 8$\times$8$\times$8 Monkhorst-Pack $\kk$-point grid.\cite{Monkhorst1976}
As expected from DFT, the band gaps are severely underestimated. 
This is corrected by applying scissor shifts to the conduction band levels in order to fit the experimental measurements for relaxed silicon. 
The same scissor shifts are kept for the strained cases. 

The phonon frequencies $\wqnu$ are obtained within density-functional perturbation theory (DFPT) using ABINIT.\cite{Gonze1997a,Gonze1997b}
The deformation potentials are considered constant, as justified later in this work.
Finally, each conduction band valley is considered separately since their degeneracy is lifted by the strain. 
The absorption coefficient is then obtained as the sum of all these separate contributions.

The density-of-states effective mass of a given band (conduction or valence) is defined as the one giving the same density of states as if the band was parabolic.
In the case of the valence band, it is obtained through\cite{Trimarchi2011,Grosso2000}
\begin{equation}
    \begin{split}
    \frac{\sqrt{\pi}}{2} \frac{1}{2\pi^2} & \left( \frac{2m_v}{\hbar^2} \right)^{3/2} (k_B T)^{3/2} = \\
    & \int_{-\infty}^{E_v} g(E) \exp\left( - \frac{E_v - E}{k_B T} \right) dE,
    \end{split}
\label{eq:holes_mass}
\end{equation}
where $E_v$ is the energy of the valence band maximum and $g(E)$ is the density of states obtained on dense meshes for each strained configuration. 
A similar expression holds for the conduction band. 
In the case of silicon, the conduction band is almost perfectly parabolic so that $m_c$ is very close to the geometric mean of the transverse and longitudinal components of the electron effective mass tensor, $m_t$ and $m_l$, respectively,\cite{Green1990}
\begin{equation}
    m_{c,i} = (m_{l,i} m_{t1,i} m_{t2,i})^{1/3},
\label{eq:elec_mass}
\end{equation}
where $i$ denotes a single valley. 
These longitudinal and transverse electron effective masses can be determined as second-order derivatives of the electronic energies, obtained from non-self-consistent computations of the band structure on dense meshes. 
In contrast, the top of the valence band does not consist of a single state and is highly anisotropic and non-parabolic so that a formula such as Eq.~\eqref{eq:elec_mass} does not hold.

\section{Indirect absorption in relaxed silicon}

In relaxed silicon, the only contribution for the absorption of infrared light ($\lambda > 750$~nm) comes from the transitions to the $\Delta$ valleys, where the conduction band minimum of silicon lies. 
The indirect band gap $E_{g,\Gamma\Delta}$ is 1.12~eV and the valley is 6-fold degenerated.
The transitions to other higher-energy valleys, such as L (see Fig.~\ref{fig:bandtransition}) or even K and U, are also responsible for visible light absorption for $E_\lambda > 2.5$~eV ($\lambda < 500$~nm). 
In this work, we focus on the region of the spectrum between 0.89 and 1.91~eV (0.65 and 1.4~$\upmu$m). 
Therefore, only the contributions from the six degenerated $\Delta$ valleys need to be included in the case of relaxed silicon, and only three gaps have to be considered in Eq.~\eqref{eq:Tsai}: the direct band gaps at $\Gamma$ and $\Delta$ ($E_{g,\Gamma}$ and $E_{g,\Delta}$, respectively) and the fundamental gap $E_{g,\Gamma\Delta}$, see Fig.~\ref{fig:bandtransition}. The experimental values are used in relaxed silicon, as explained in our methodology. 

Different phonon modes $\nu$ have to be considered for the intervalley transitions. In silicon, 6 modes are present and should be taken into account. Following Ridley,\cite{Ridley1990} Tsai included only the transverse optical (TO) and acoustic (TA) modes.\cite{Tsai2018} However, it has recently been shown by Vandenberghe~\textit{et al.}~that the deformation potential for the longitudinal optical (LO) mode is far from being negligible,\cite{Vandenberghe2015} while the longitudinal acoustic (LA) mode can be safely ignored.\cite{Vandenberghe2015} Therefore, we also include the contribution of the LO mode and use these deformation potentials, that are given in Table~\ref{tab:ModelParameter}. The values used by Tsai are given in brackets for comparison. 
\begin{table}
\caption{Physical parameters used to compute the indirect absorption coefficient of relaxed silicon. The values used by Tsai that differ from our work are given in brackets.~\cite{Tsai2018}}
    \begin{ruledtabular}
    \begin{tabular}{cc}
    Parameters & Value \\ 
    \hline
    $n_r$ & 3.42 \\
    $\rho_L$ & 2.328 g/cm$^3$ \\
    $m_c$ & 0.36 (0.328)\,$m_0$ \\ 
    $m_v$ & 0.81 (0.55)\,$m_0$ \\
    T & 300\,K \\
    $E_p$ & 12 (25) eV \\
    $\Gamma_d$ & 1.35 eV \\
    $E_{g,\Gamma}$ & 3.2 eV \\
    $E_{g,\Gamma\Delta}$ & 1.12 (1.1557) eV \\
    $E_{g,\Delta}$ & 4 eV \\
    $D_\mathrm{TA}$ & $4.1 \times 10^8$ ($8.8 \times 10^8$) eV/cm \\ 
    $D_\mathrm{TO}$ & $1.2 \times 10^9$ ($1.3 \times 10^9$) eV/cm \\ 
    $D_\mathrm{LO}$ & $2.2 \times 10^9$ ($0$) eV/cm \\ 
    \end{tabular}
    \end{ruledtabular}
\label{tab:ModelParameter}
\end{table}

We obtain the conduction and valence bands density-of-states effective masses using the methodology described in the previous section and obtain different values as the ones used by Tsai, see Table~\ref{tab:ModelParameter}.
Particularly, the valence band effective mass is $47$\% larger but in better agreement with experimental evidence.\cite{Barber1967}
The conduction band effective mass is also in good agreement with previous results.\cite{Ramos2001,Bouhassoune2010,Barber1967}
The parameter related to the electron-photon interactions is also adapted to better fit the experimental absorption coefficient.
All the other parameters required by the indirect-light absorption model are given in Table~\ref{tab:ModelParameter}. 

\begin{figure}
    \includegraphics[width=1\linewidth]{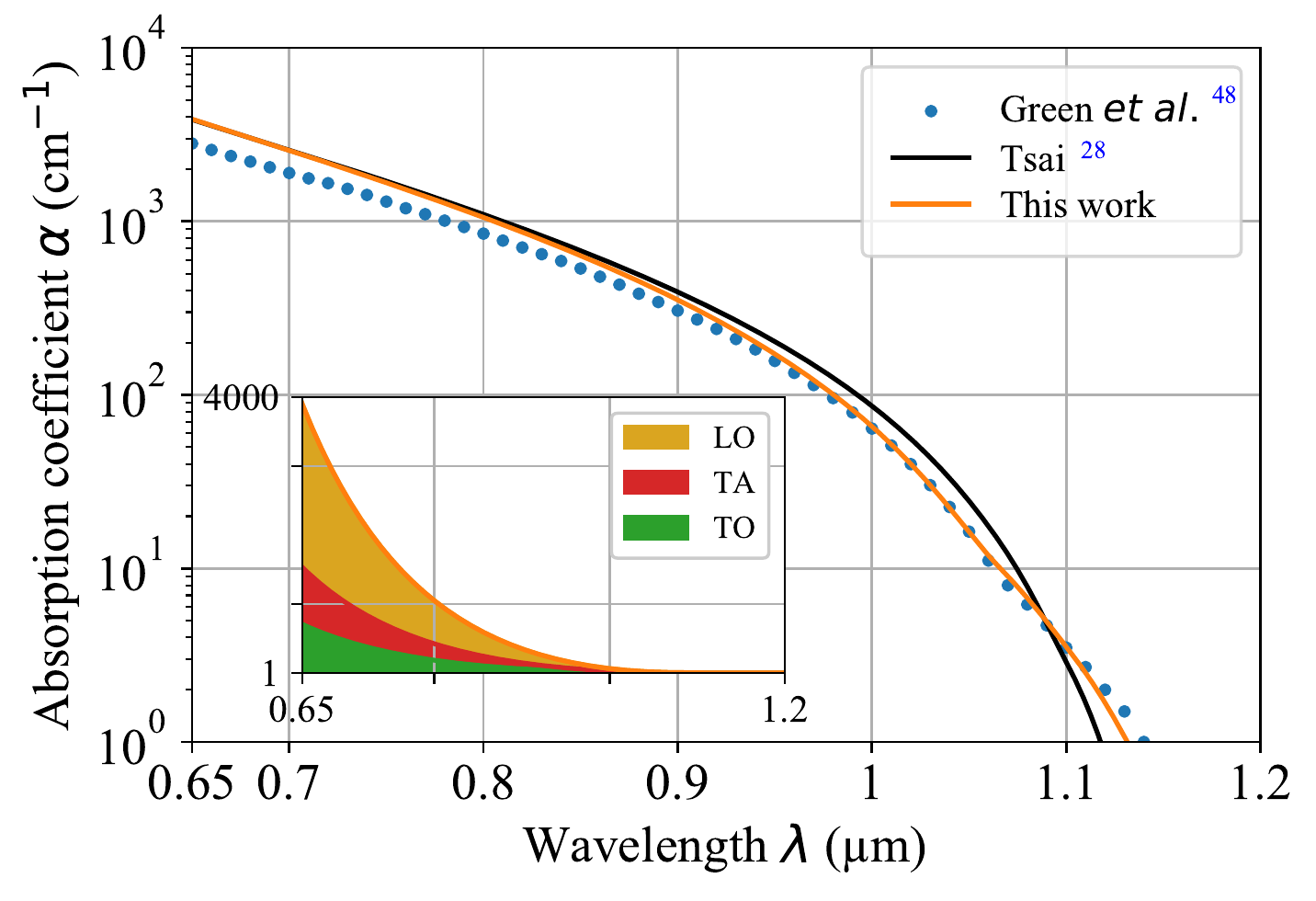}
    \caption{Comparison between our indirect absorption coefficient computed including the contributions of the TA, TO and LO modes (solid orange line) with that obtained by Tsai, including only the TO and TA modes with different parameters (solid black line). These are compared with the absorption coefficient measured experimentally by Green~\textit{et al.}~(blue circles).\cite{Green1995} 
    The inset shows the contributions of the TA, TO and LO modes to the absorption coefficient.}
\label{fig:TsaiIllustration}
\end{figure}

Figure~\ref{fig:TsaiIllustration} compares the absorption coefficient obtained by Tsai (including only the TO and TA modes, and using the parameters in brackets in Table~\ref{tab:ModelParameter}) with our result obtained by taking into account the LO mode and the deformation potentials from Vandenberghe~\textit{et al.},\cite{Vandenberghe2015} evaluated at the $\qq$ point connecting the valence band maximum to the conduction band minimum ($\qq = \Delta$), and our computed effective masses.
The contributions of the different modes are further analyzed in the inset of Fig.~\ref{fig:TsaiIllustration}. 
In order to evaluate this model, the data from Green~\textit{et al.}\cite{Green1995}~for the absorption coefficient of silicon is also represented in Fig.~\ref{fig:TsaiIllustration}. 
Due to its availability, this data set is usually taken as a reference for silicon absorption to compare and check the validity of different absorption models.
It can be seen from Fig.~\ref{fig:TsaiIllustration} that the agreement between Tsai results, ours, and the experimental data is very good on a large portion of the spectrum. 
However, close to the cutoff wavelength of 1.14~$\upmu$m, our model allows for a better description of the shape of the absorption spectrum, in contrast with the case where the LO mode is not taken into account. 
Indeed, this mode is dominating the absorption coefficient spectrum and should not be neglected, particularly in the region of interest for near-infrared applications.

In the following, we extend this model to take into account the impact of strain on the absorption of silicon using DFT and DFPT to compute the evolution of important parameters such as the effective masses, the band gaps and the phonons energies. This allows us to efficiently compute and predict the optical properties of highly-strained silicon.

\section{Highly strained silicon}
\label{sec:results}

\subsection{Electronic band structure}
\label{sec:bandstructure}

In the energy region of interest, i.e., 0.89--1.91~eV, only the indirect absorption through the $\Delta$ valleys have to be considered in the case of relaxed silicon. 
For strained-silicon, we also need to monitor the indirect band gaps between the valence band maximum at $\Gamma$ and other valleys of the conduction band, such as L, K, or U ($E_{g,\Gamma \LL}$, $E_{g,\Gamma \mathrm{U}}$ and $E_{g,\Gamma \mathrm{K}}$, respectively). 
Indeed, even if these indirect gaps are larger than 1.91~eV at zero strain, they could decrease and lead to important contributions to the absorption spectrum of highly strained silicon. 
Table~\ref{tab:bandresults} gives the band gaps obtained with DFT, together with the experimental values and the applied scissor shifts. 
\begin{table}[hptb!]
    \caption{Relevant band gaps for the indirect absorption coefficient of silicon, obtained with DFT. Experimental data are also provided, together with the applied scissor shifts. All the values are expressed in eV.}
    \begin{ruledtabular}
    \begin{tabular}{cccc}
         & DFT & Expts.\cite{Kane1966} & Scissor \\
        \hline
        $E_{g,\Gamma\Delta}$ & 0.61 & 1.12 & 0.51 \\
        $E_{g,\Gamma}$       & 2.55 & 3.2  & 0.65 \\
        $E_{g,\Delta}$       & 3.28 & 4    & 0.72 \\
        $E_{g,\Gamma \LL}$     & 1.46 & 2.5  & 1.04 \\
        $E_{g,\Gamma \mathrm{U}}$     & 1.24 &  -    & 1.04 \\
        $E_{g,\Gamma \mathrm{K}}$     & 1.24 &  -    & 1.04 \\
    \end{tabular}
    \end{ruledtabular}
\label{tab:bandresults}
\end{table}
The scissor shifts are different for each transition, because the underestimation of the gap is not uniform.
In general, the shift of the conduction band energies increases with the electron energy.\cite{Hybertsen1986} 
In this work, we fix the scissor shifts depending on the experimental values of the gaps in relaxed silicon, and keep the same values for the strained cases.
Since the L, U, and K valleys have similar energies, we apply the same scissor shift to the corresponding gaps. 

\begin{figure}
    \includegraphics[width=1\linewidth]{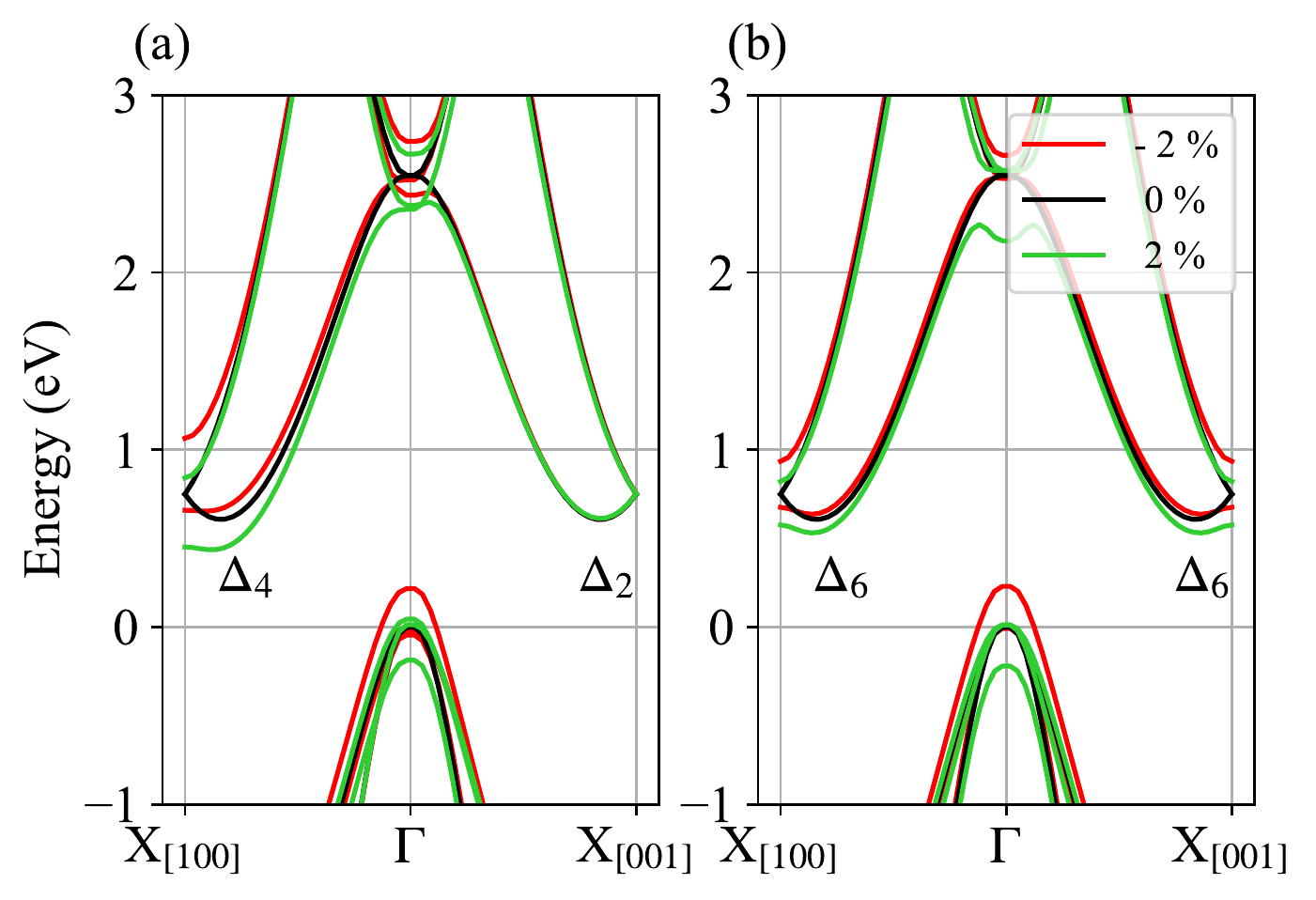}
    \caption{Electronic DFT band structure of relaxed (black) and highly strained silicon, along different $\Gamma-$X directions. 
    A uniaxial strain of $-2$\% (red) and $+2$\% (green) is applied along the (a) [110] and (b) [111] directions.
    The zero of energy is set at the valence band maximum in the relaxed case.}
\label{fig:bandstructure}
\end{figure}

When a uniaxial strain along the [110] direction is applied, some crystal symmetries are broken which lifts the degeneracy of the $\Delta$ valleys into the $\Delta_2$ (including the two valleys along [001]) and the $\Delta_4$ (including the four valleys along [100] and [010]) valleys, as shown in Fig.~\ref{fig:bandstructure}(a).
The L valleys, initially 4-fold degenerated, are split into two groups: the ones located along [111] and [11$\bar{1}$] ($\LL_{2a}$), and the two others ($\LL_{2b}$). 
A similar analysis can be done for the K valleys that lead to three different pockets ($\KK_a$, $\KK_b$, and $\KK_c$) when strain is applied. 
Finally, the analysis for the U valleys is the same as the ones related to the K valleys and is not repeated here. 

\begin{figure}
    \includegraphics[width=1\linewidth]{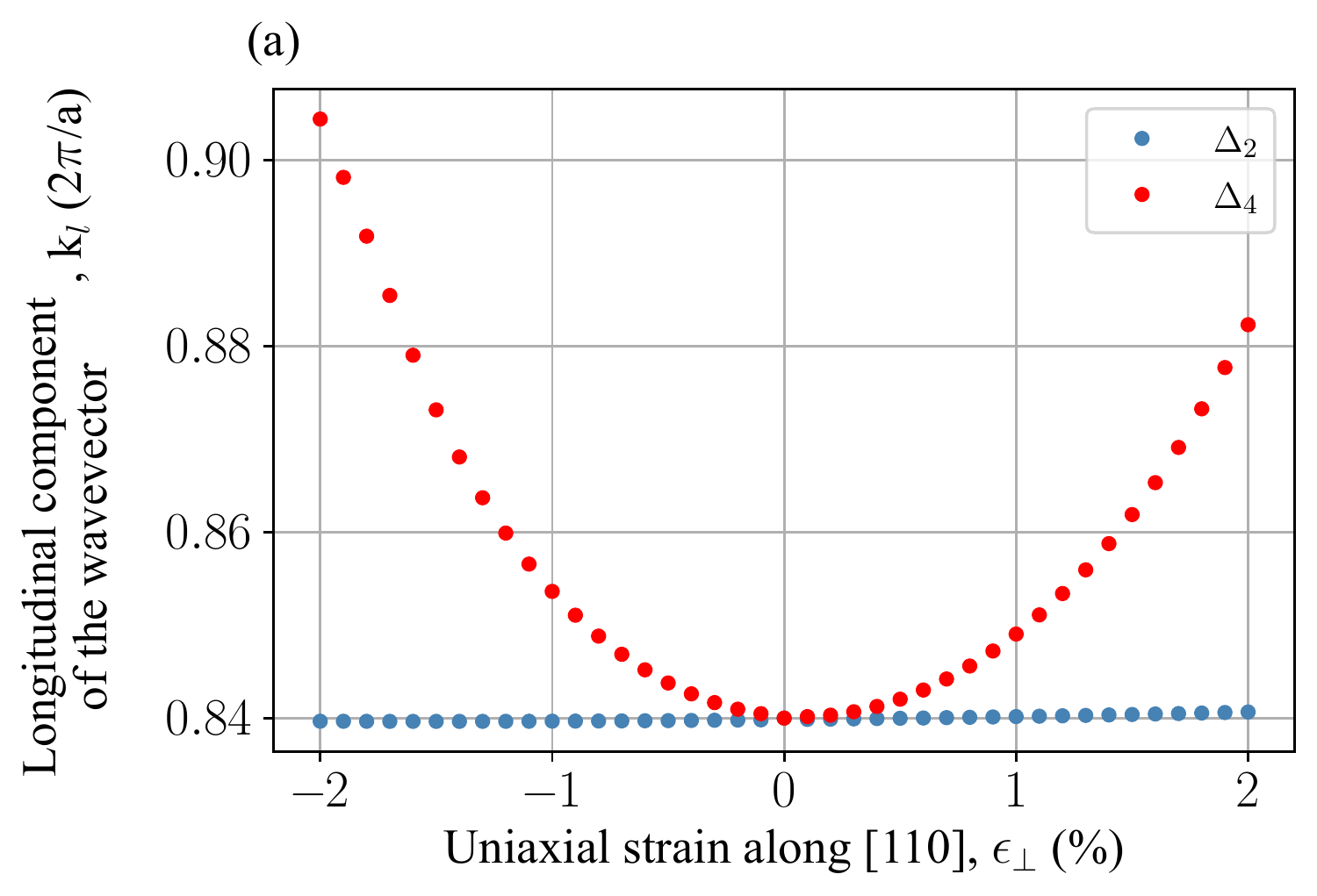}
    \includegraphics[width=1\linewidth]{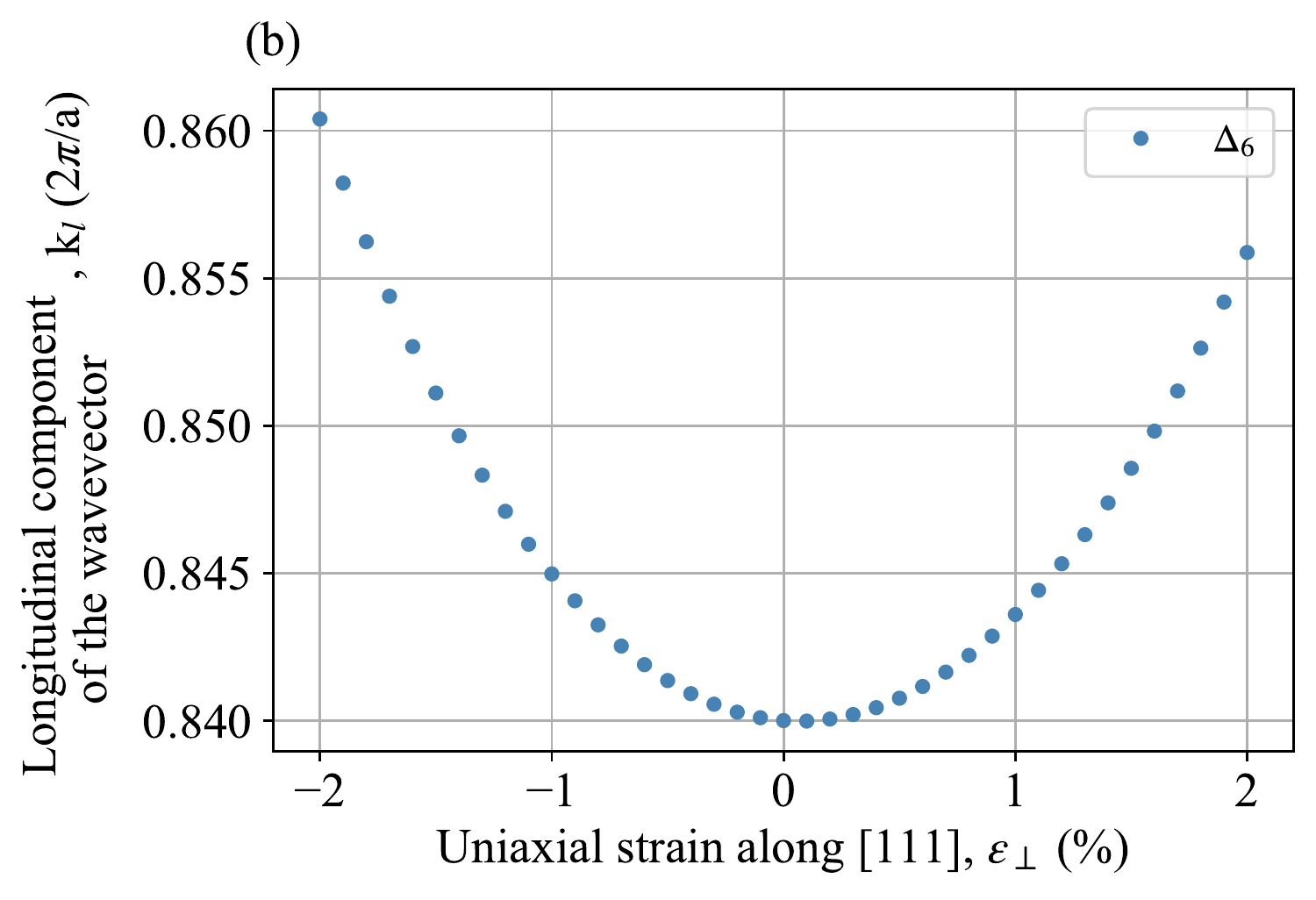}
    \caption{Evolution of the longitudinal component of the $\Delta$-valley wavevector (in units of $2\pi/a$, with $a$ the unit cell parameter) with a uniaxial strain applied along the (a) [110] and (b) [111] directions.}
\label{fig:valleydisplacement}
\end{figure}

Before determining the different band gaps and effective masses for each strained structure, we need to precisely calculate the position of the different $\Delta$ points when strain is applied.\cite{Bouhassoune2010,Bouhassoune2015}
Using very dense grids in reciprocal space around these valleys, we found that the displacement of the $\Delta$ points occurs mainly along the longitudinal direction. 
In the case of a uniaxial strain in the [110] direction, the $\Delta_2$ valleys show almost no displacement while the $\Delta_4$ valleys show a quadratic displacement, as shown in Fig.~\ref{fig:valleydisplacement}(a). 
These displacements are taken into account in the following.

\begin{figure}
    \includegraphics[width=1\linewidth]{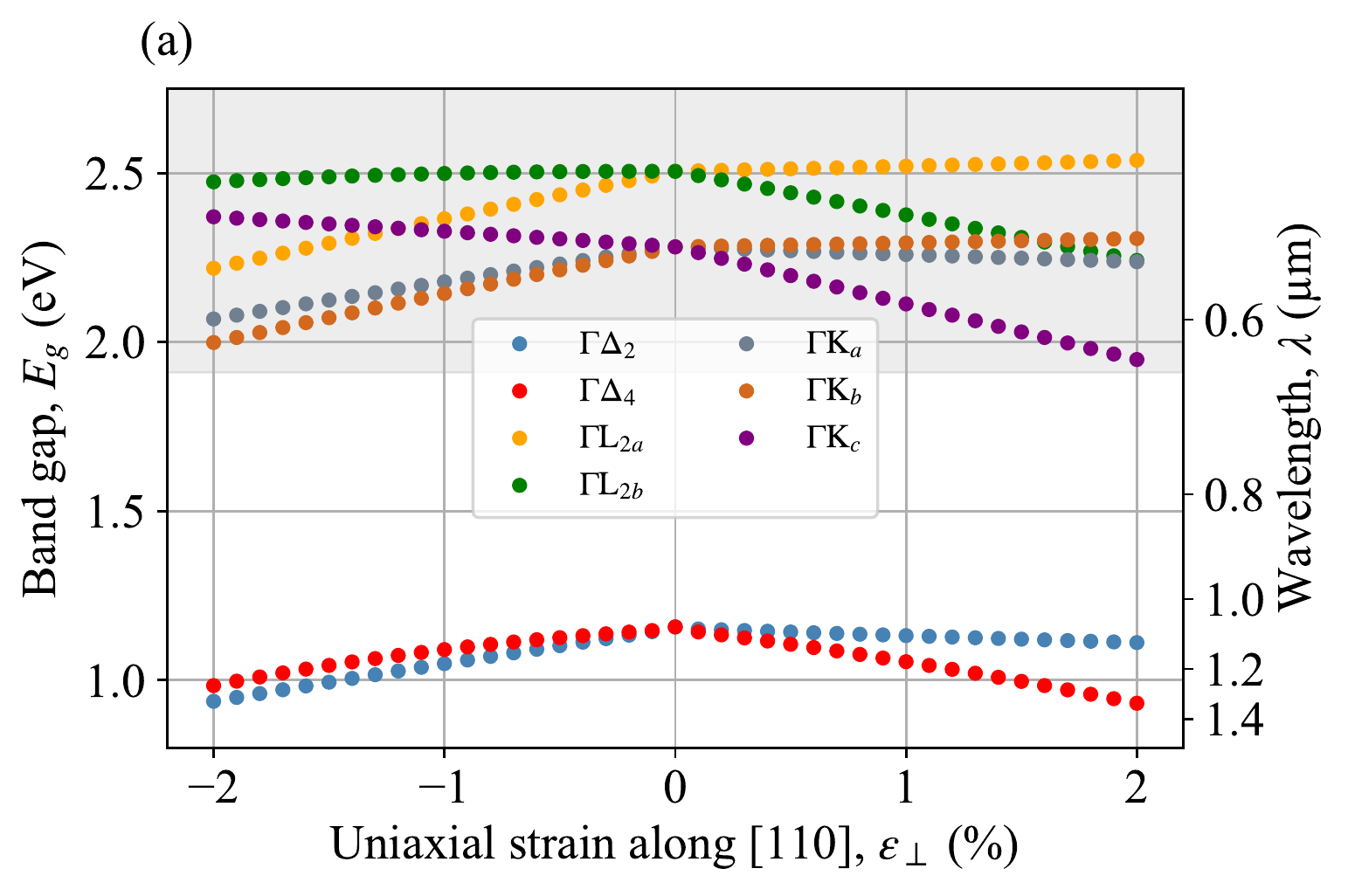}
    \includegraphics[width=1\linewidth]{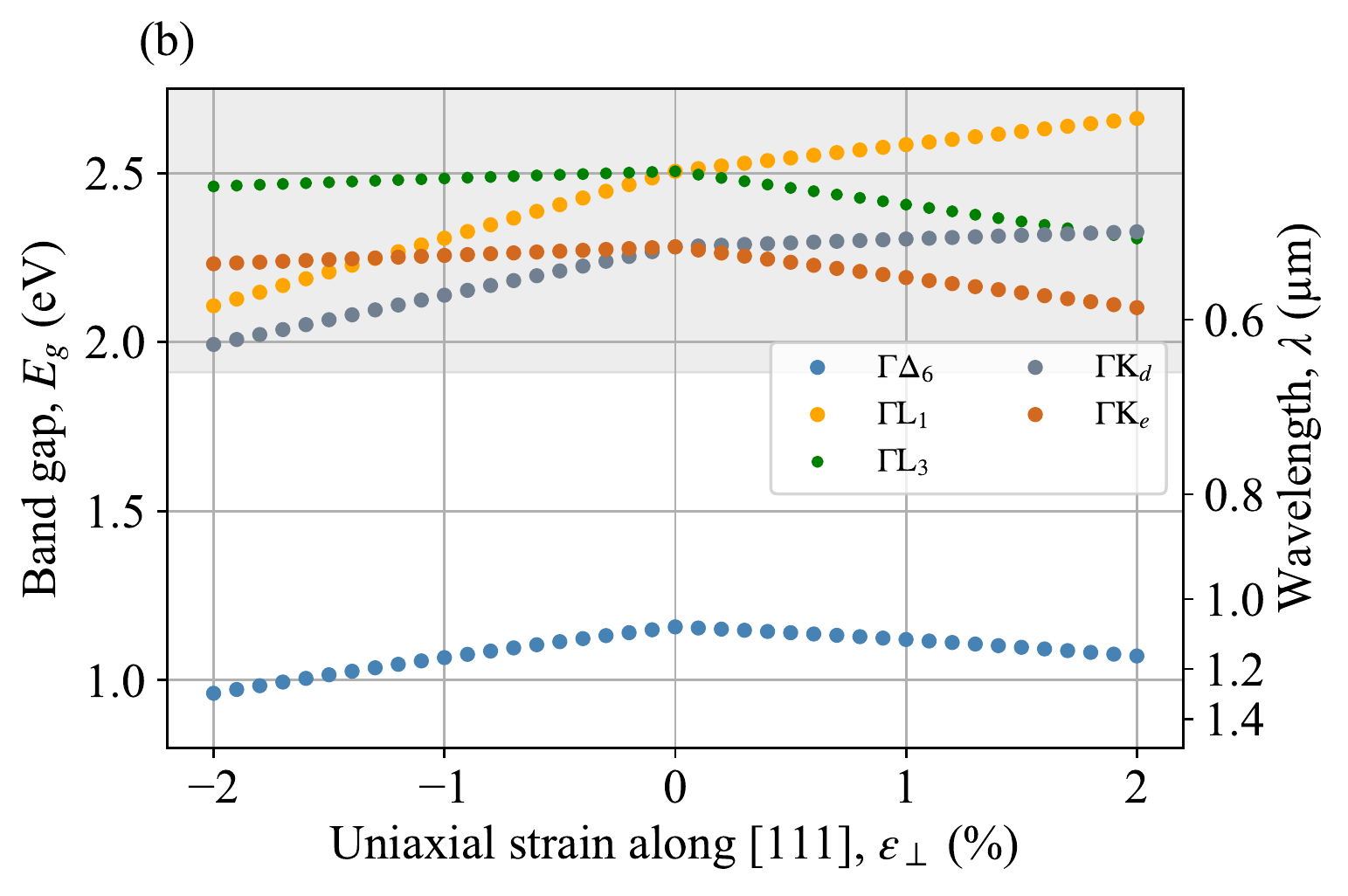}
    \caption{Evolution of the indirect band gaps of silicon with a uniaxial strain applied along the (a) [110] and (b) [111] directions. The gray shaded zone denotes energies outside the region of interest.}
\label{fig:indirectbandgap}
\end{figure}

The evolution of the band gaps with the uniaxial strain applied along [110] is represented in Fig.~\ref{fig:indirectbandgap}(a). 
The fundamental band gap reduces significantly under both tensile (positive) and compressive (negative) strains, with a sensitivity of $-0.11$~eV/\% in both cases.
However, the indirect band gaps related to the K, U and L valleys all remain larger than 1.91~eV with a maximum strain level of 2\%. 
Therefore, these will not lead to additional contributions to the absorption coefficient in the region of interest, and only the $\Delta$ valleys should be considered. 

In the case of a uniaxial strain along the [111] direction, the 6-fold degeneracy of the $\Delta$ valleys remains, see Fig.~\ref{fig:bandstructure}(b),
while the L valley along [111] ($\LL_{1}$) is separated from the three others ($\LL_{3}$).
For the K and U valleys, there are two different band gaps, namely $E_{g,\Gamma \KK_d}$ and $E_{g,\Gamma \KK_e}$.
The displacement of the $\Delta_6$ valleys with the applied strain is smaller in this case, see Fig.~\ref{fig:valleydisplacement}(b). 
The evolution of the band gaps with the uniaxial strain along the [111] direction is represented in Fig.~\ref{fig:indirectbandgap}(b). 
The fundamental band gap also reduces, with a sensitivity of $-0.10$~eV/\% ($-0.05$~eV/\%) for compressive (tensile) strain. 
The higher-energy $E_{g,\Gamma \LL}$ and $E_{g,\Gamma \KK}$ band gaps decrease rapidly with the strain level, but still remain above the considered spectrum. Thus, they will not impact our computed absorption coefficient. 
Again, only the $\Delta$ valleys should be considered.

\subsection{Effective masses}
\label{sec:eff_mass} 

\begin{figure}
    \includegraphics[width=1\linewidth]{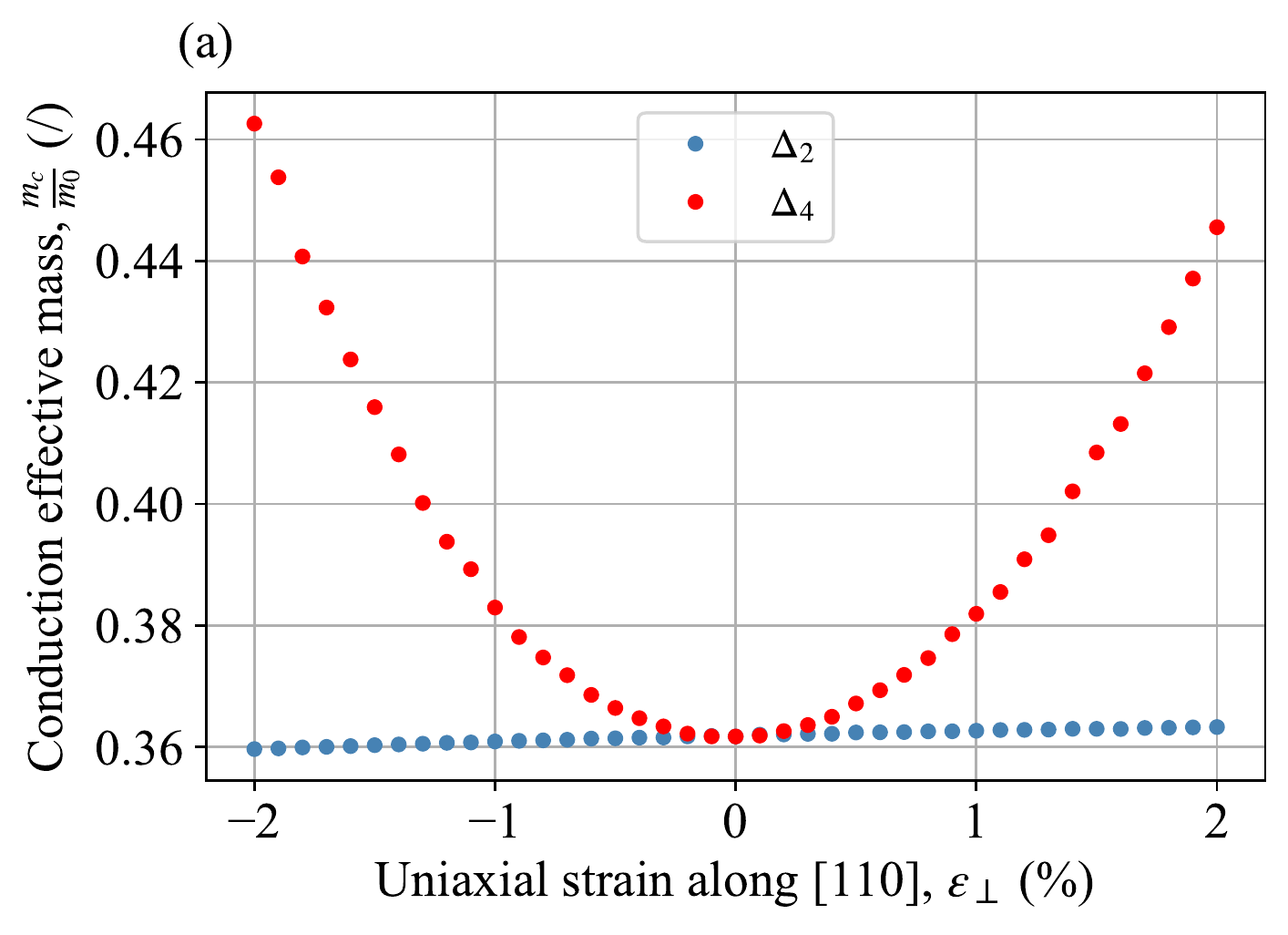}
    \includegraphics[width=1\linewidth]{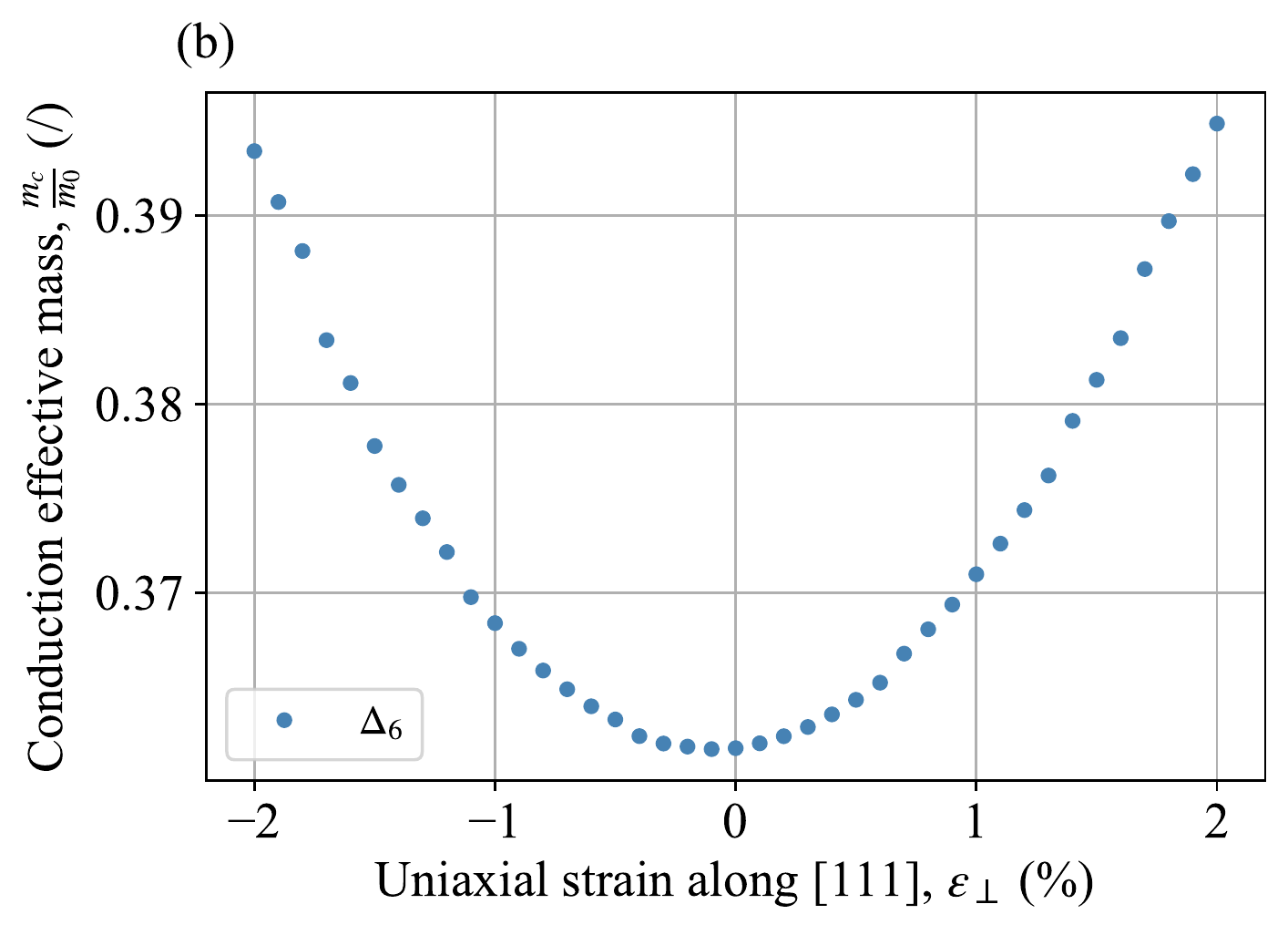}
    \caption{Evolution of the conduction band density-of-states effective masses with a uniaxial strain applied along the (a) [110] and (b) [111] directions.}
\label{fig:electronmass}
\end{figure}

Figure~\ref{fig:electronmass} shows the evolution of the conduction band density-of-states effective masses for the different $\Delta$ valleys with the applied uniaxial strain. 
It can be seen that the strain levels considered in this work are not sufficient to affect significantly the conduction band effective masses. 
As shown later, this effect only slightly impacts the absorption coefficient of highly strained silicon. It can thus safely be ignored. 

\begin{figure}
    \includegraphics[width=1\linewidth]{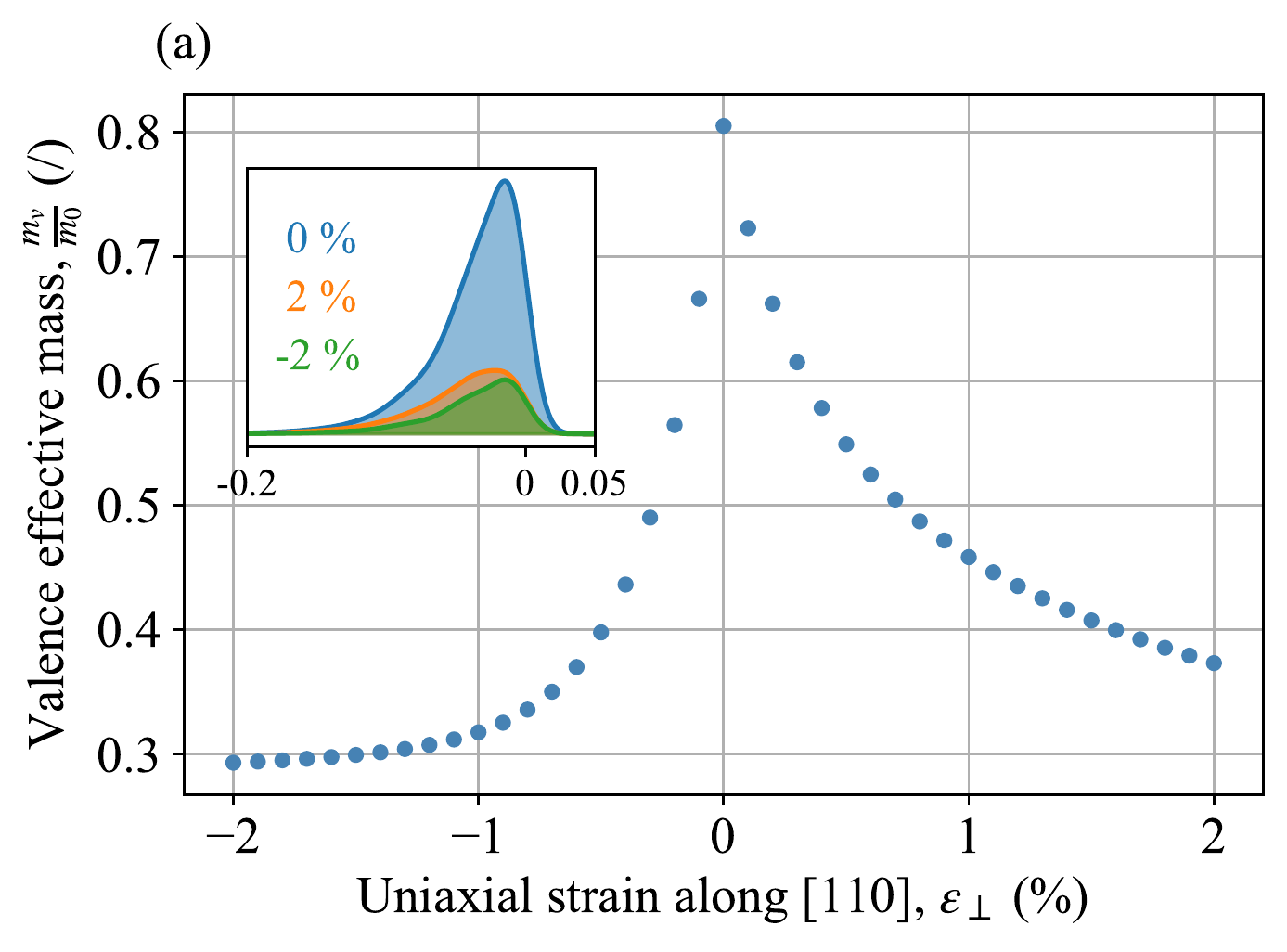}
    \includegraphics[width=1\linewidth]{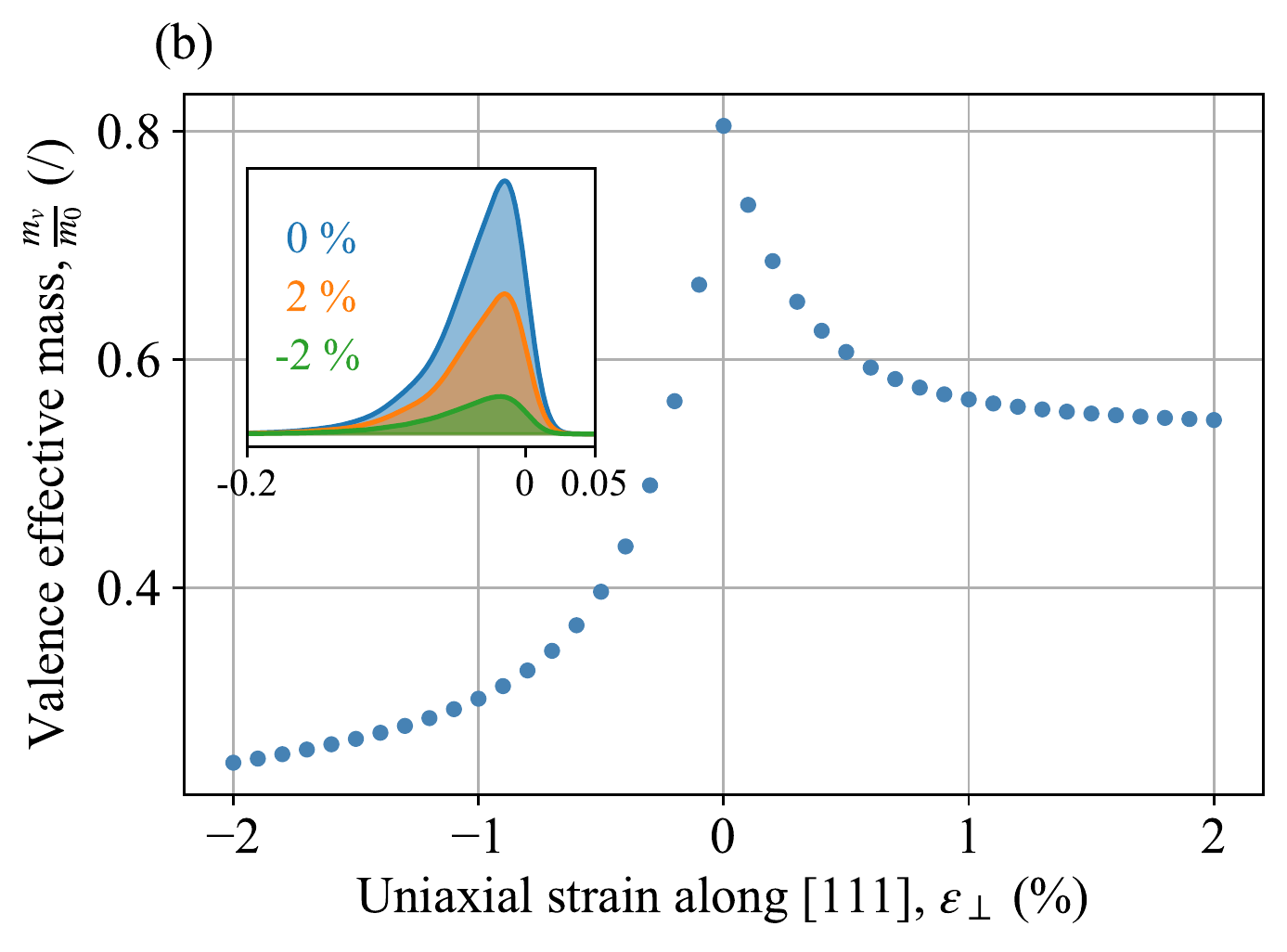}
    \caption{Evolution of the valence band density-of-states effective mass with a uniaxial strain applied along the (a) [110] and (b) [111] directions.}
\label{fig:holesmass}
\end{figure}

Figure~\ref{fig:holesmass} shows the evolution of the valence band density-of-states effective mass with the applied uniaxial strain. 
In contrast with the conduction case, the valence band effective mass decreases significantly with the applied strain. 
This effect is precisely due to the large anisotropy and non-parabolicity of the valence bands, that become more parabolic as strain is applied. 
The density of states is largely affected, as shown in the insets of Fig.~\ref{fig:holesmass}, depicting the evolution of the integrand on the right-hand side of Eq.~\eqref{eq:holes_mass} close to the valence band maximum.

\subsection{Phonons}
\label{sec:phonons}

\begin{figure}[hpbt!]
    \includegraphics[width=1\linewidth]{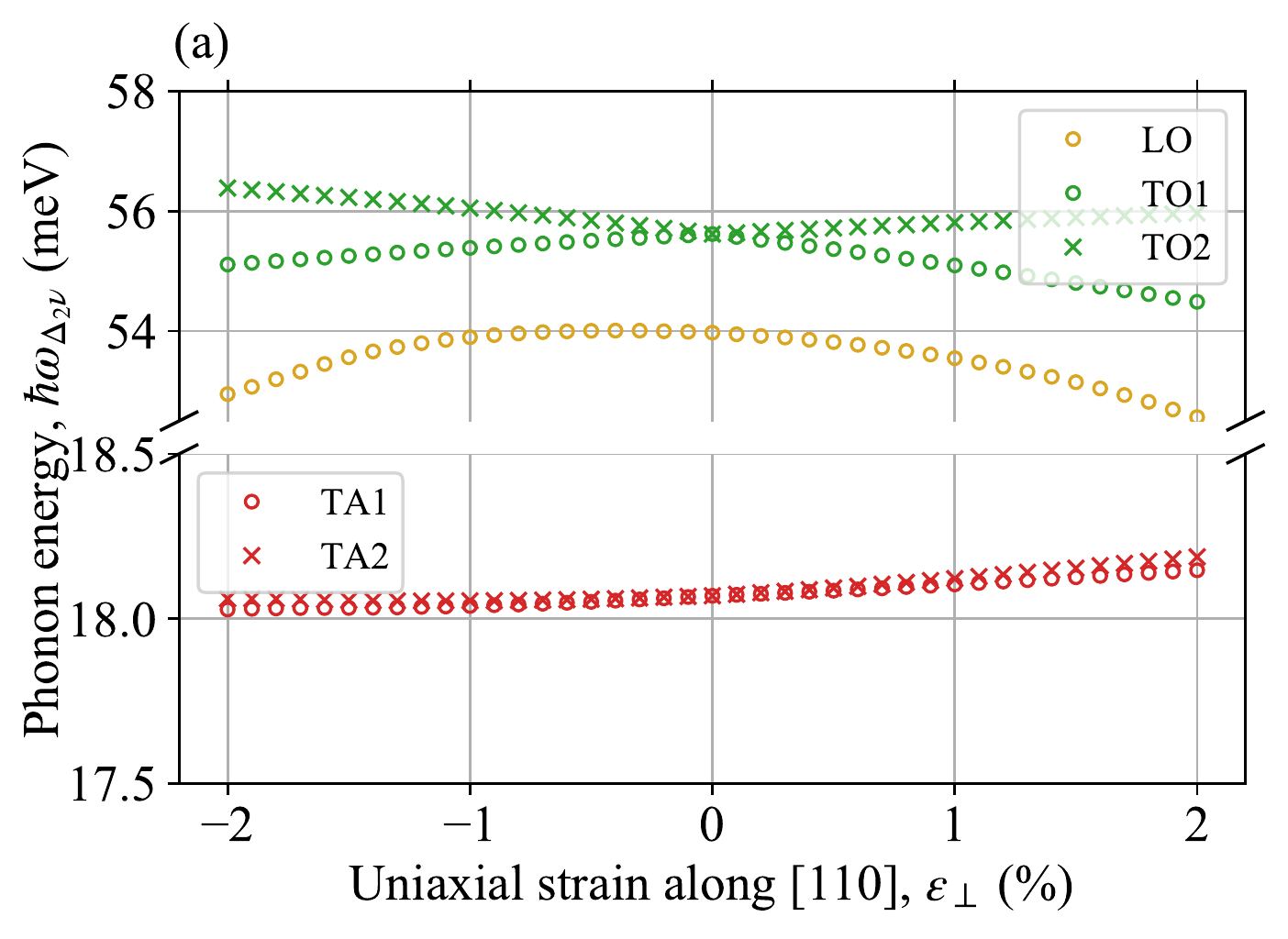}
    \includegraphics[width=1\linewidth]{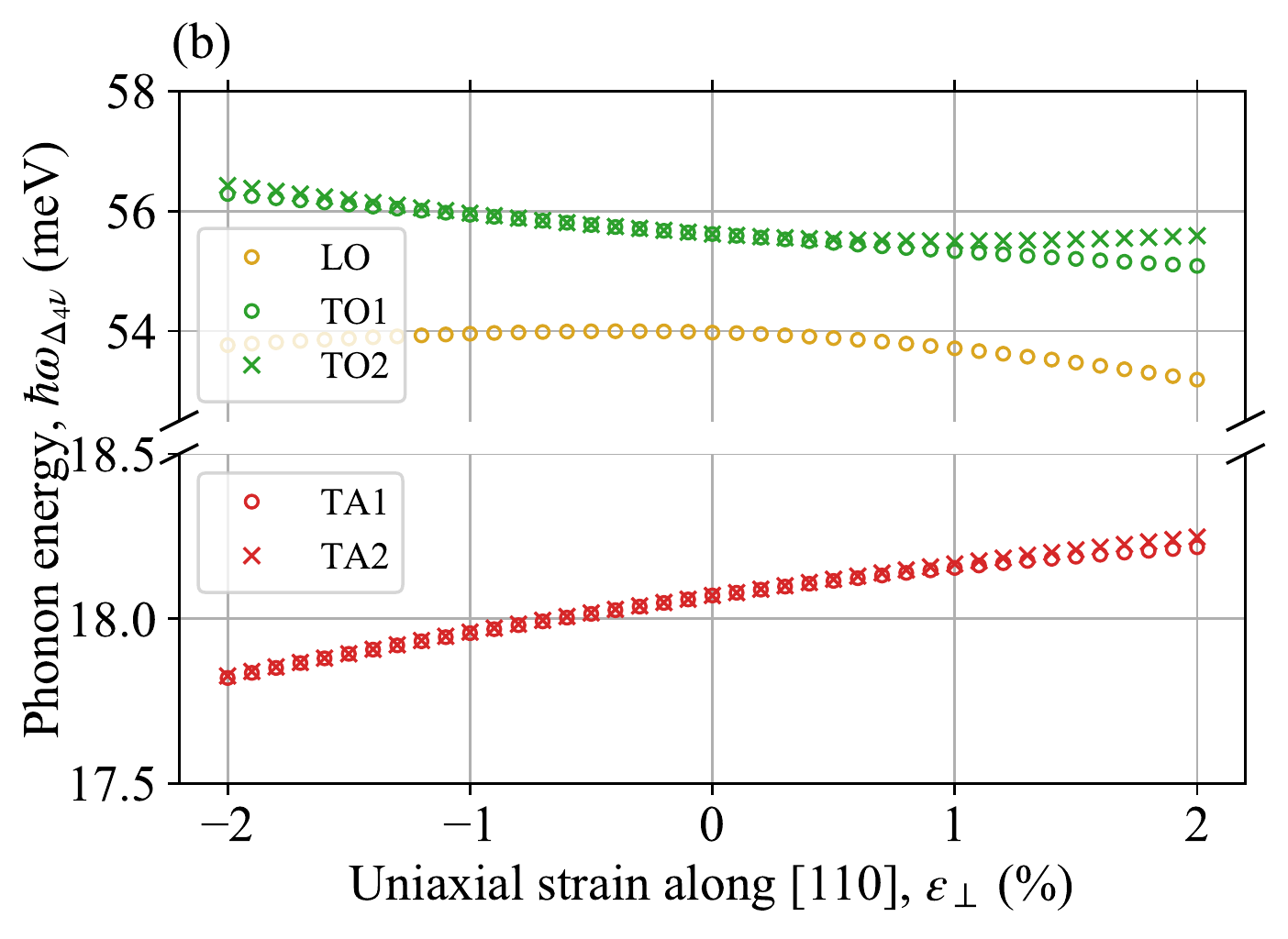}
    \includegraphics[width=1\linewidth]{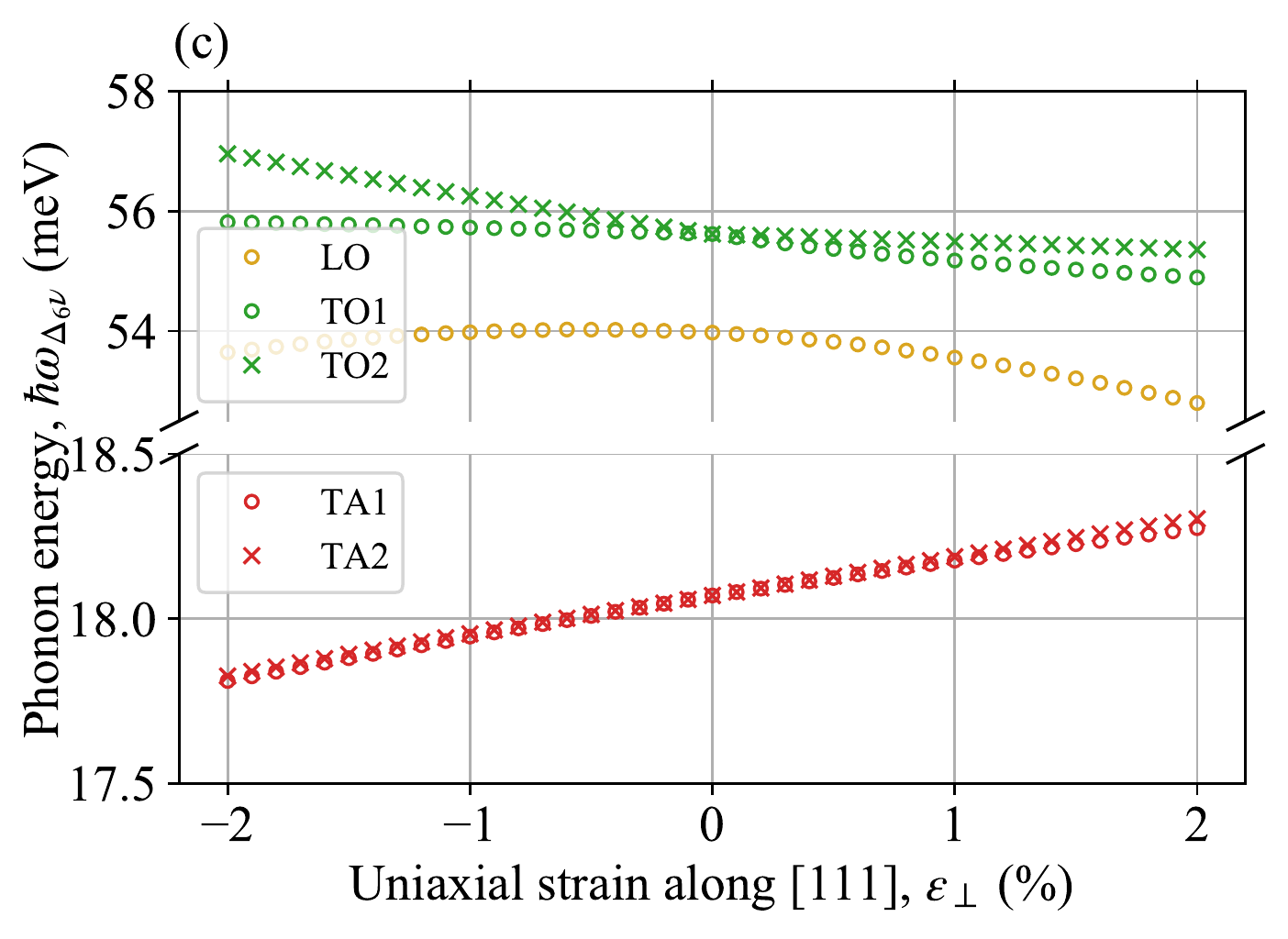}
    \caption{Evolution of the optical and acoustic phonons energies at (a) the $\Delta_2$ and (b) the $\Delta_4$ valleys under uniaxial [110] strain, and (c) the $\Delta_6$ valleys under uniaxial [111] strain.}
\label{fig:phonons}
\end{figure}

The impact of strain on the phonon frequencies must also be carefully analyzed.
As explained in Sec.~\ref{sec:model}, only the phonons with a wavevector $\qq$ connecting the valence band maximum at $\Gamma$ to the conduction band minimum at $\Delta$ should be considered. 
It has recently been shown that the TO, TA and LO modes show a strong electron-phonon coupling for such interband transitions, while the LA mode plays a negligible role.\cite{Vandenberghe2015}
Here, we analyze the behavior of the phonon frequencies for these modes and wavevector $\qq = \Delta$. 
Figure~\ref{fig:phonons} shows the evolution of these $\omega_{\Delta\nu}$ with the different types of strain considered, as computed with DFPT. 
The changes in phonon frequencies are overall negligible, and these changes only have a small impact on the absorption coefficient of highly strained silicon.

The deformation potentials $\Dqnu$ being obtained from the electron-phonon coupling matrix elements,\cite{Vandenberghe2015} one can make the following assumption. 
Most of the changes of the electronic structure of silicon due to the strain are related to the value of the band gap, while the character of the conduction band minimum and valence band maximum does not strongly vary.
Changes in the band gap do not have a large impact on the electron-phonon coupling matrix elements.
The phonon properties are also quite robust with respect to the effect of strain, at least for the strain levels considered in this work. 
Indeed, the structure only changes by a few percent.
For these reasons, we can consider the deformation potentials to be constant with respect to the strain level, as long as the latter remains sufficiently small. 

\subsection{Indirect absorption}
\label{sec:ind_abs}

\begin{figure}
    \includegraphics[width=1\linewidth]{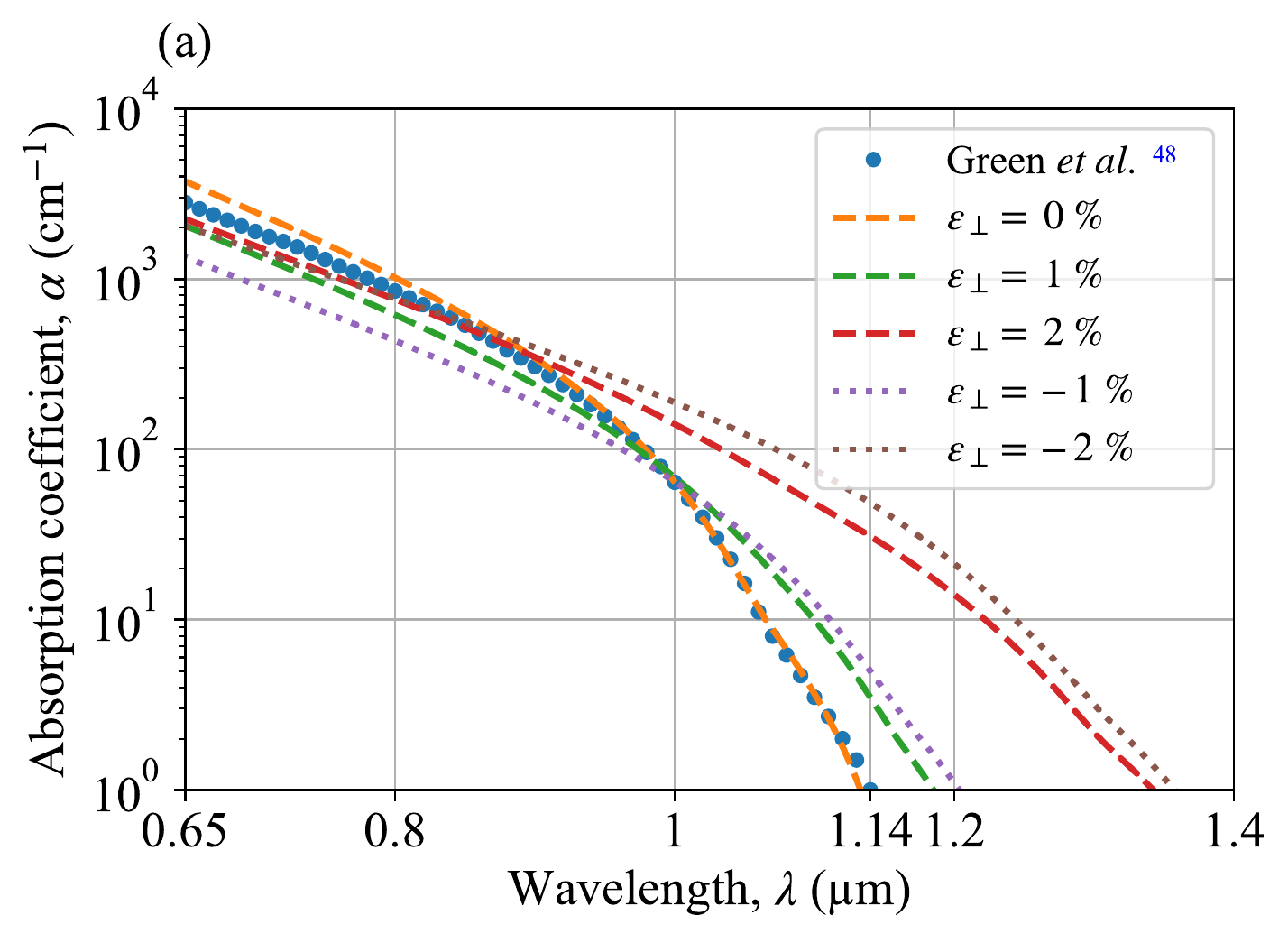}
    \includegraphics[width=1\linewidth]{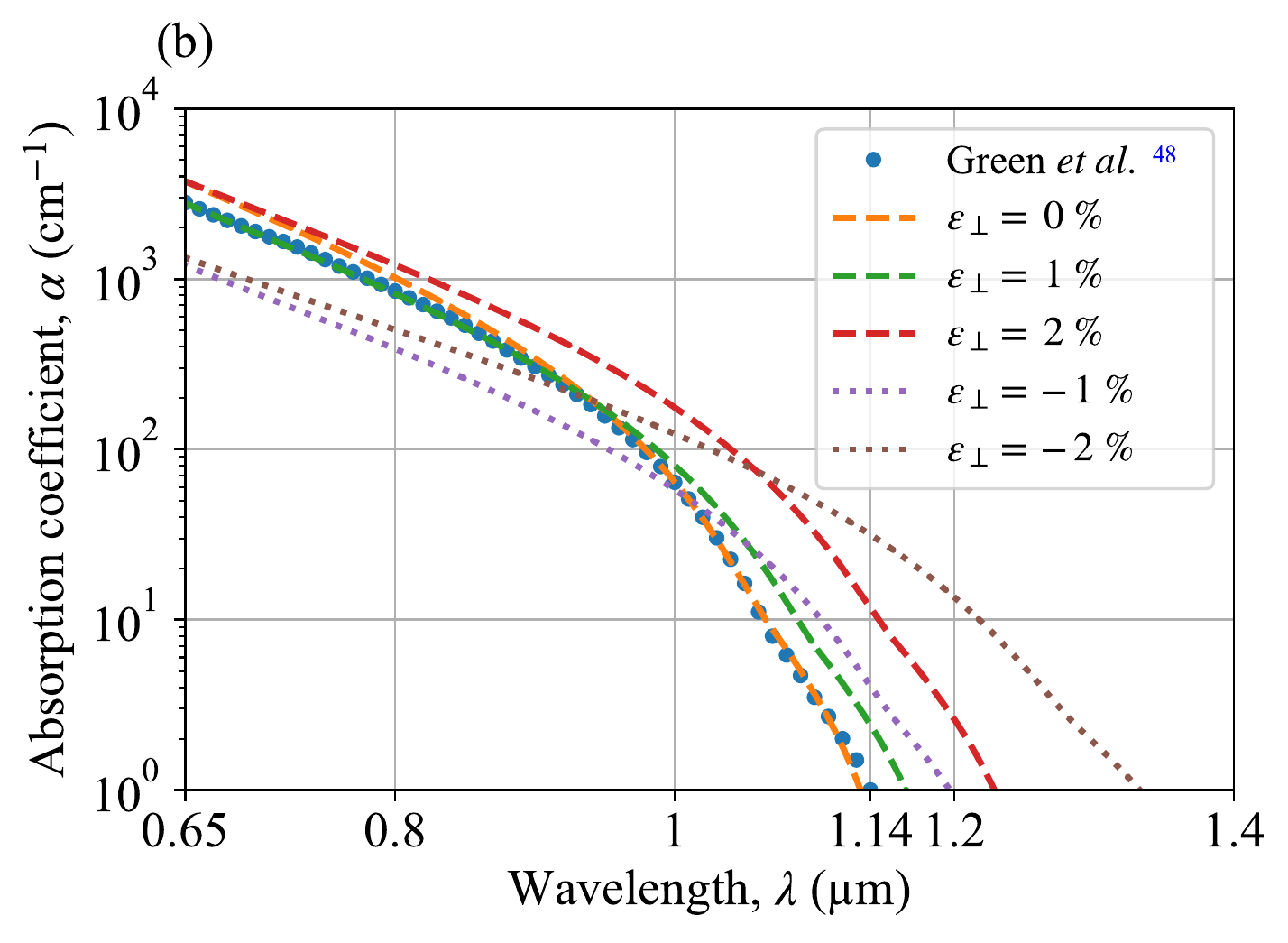}
    \caption{Absorption coefficient spectrum of silicon for different uniaxial strain levels in the (a) [110] and (b) [111] directions.}
\label{fig:absorption}
\end{figure}

The absorption coefficient of highly strained silicon is shown in Fig.~\ref{fig:absorption}. 
Each $\Delta$ valley is considered separately, and their contributions are added up following Eq.~\eqref{eq:alpha_tot} and considering the changes in band gaps, effective masses and phonon frequencies due to the strain.
Applying uniaxial compressive strain along the [111] direction or tensile and compressive strain along the [110] direction lead to very similar performances. 
On the other hand, a tensile strain along [111] leads to poorer results, which is related to a smaller reduction of the fundamental band gap, see Fig.~\ref{fig:indirectbandgap}.

\begin{figure}
    \includegraphics[width=1\linewidth]{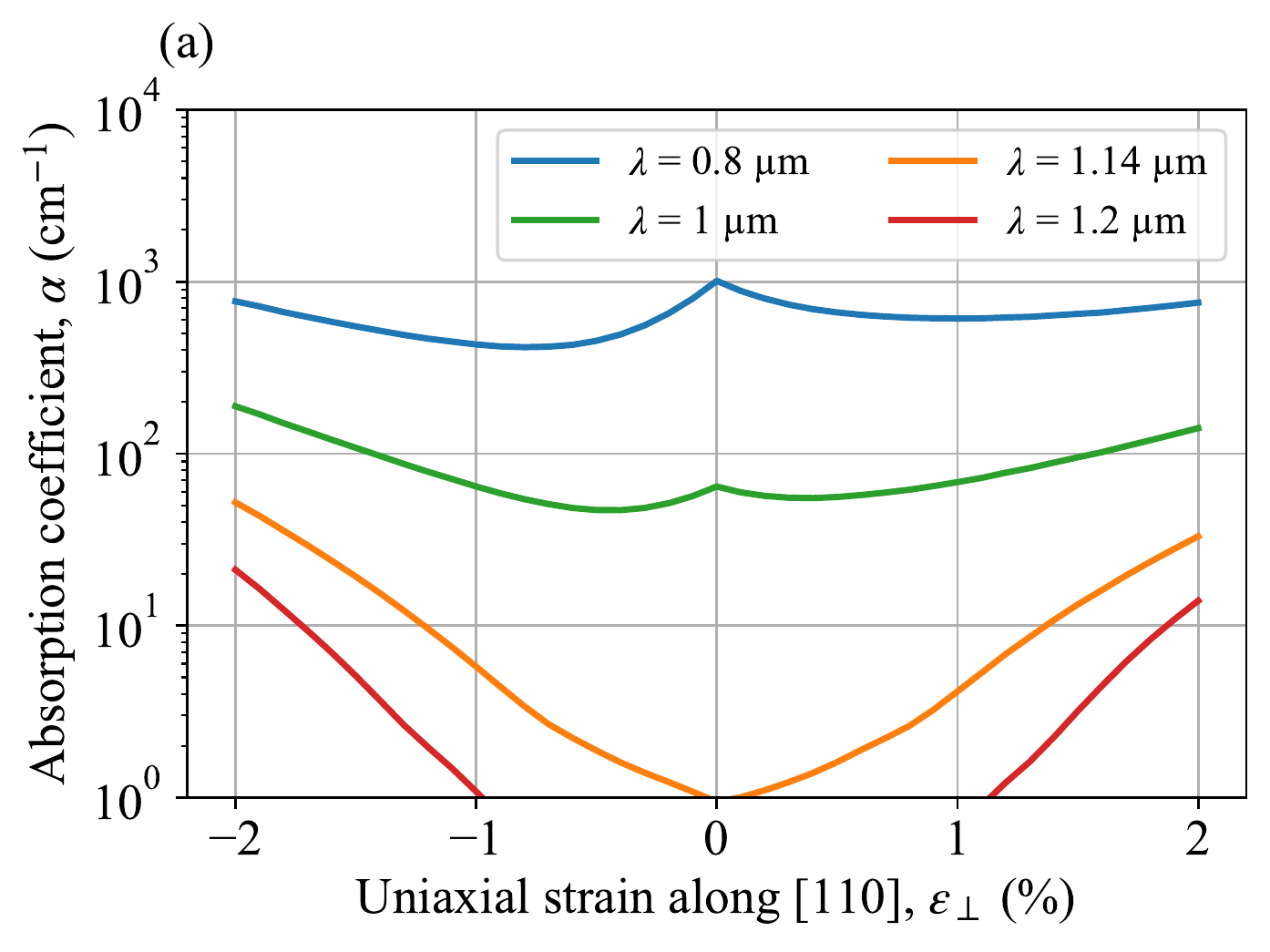}
    \includegraphics[width=1\linewidth]{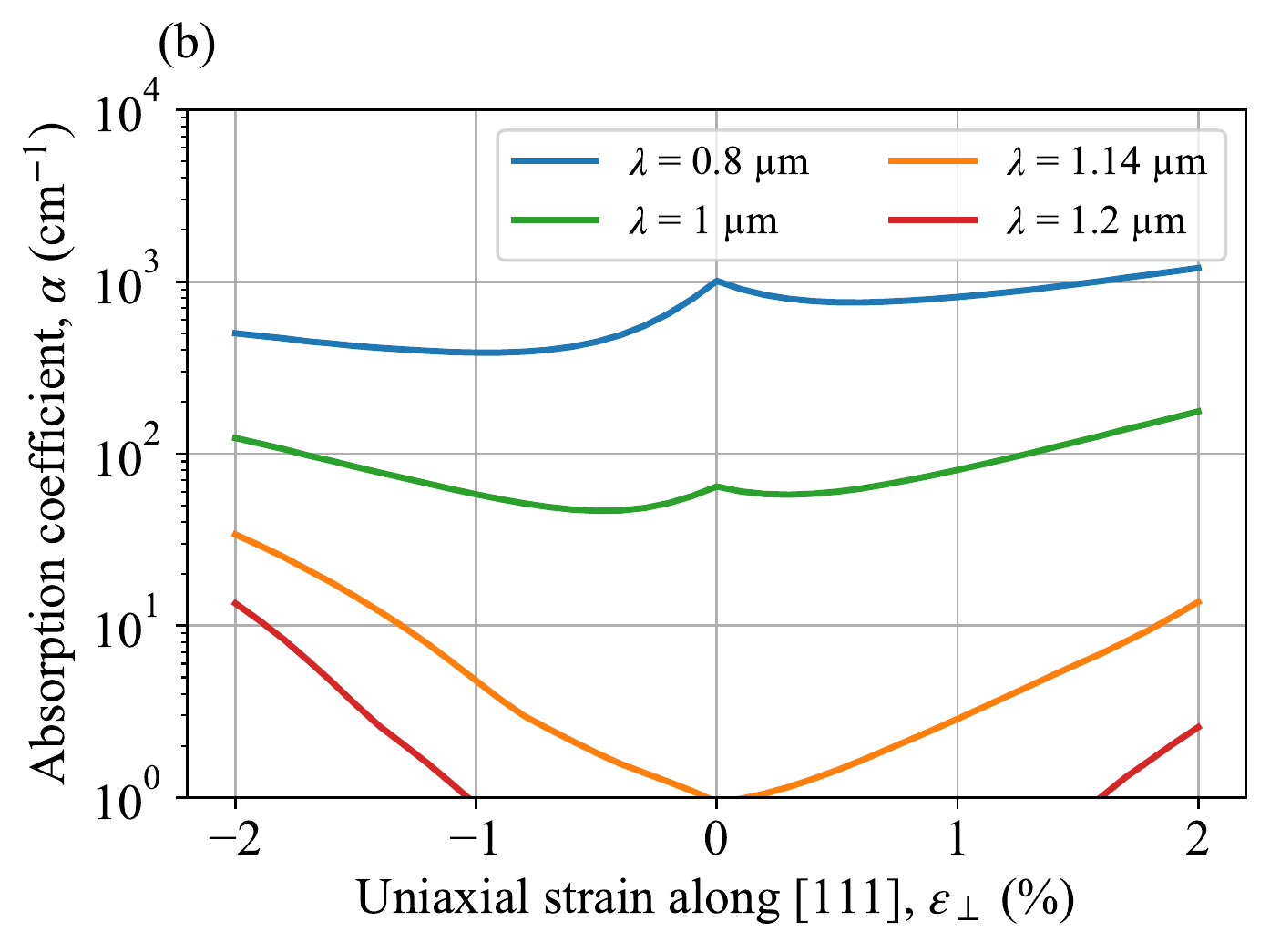}
    \caption{Evolution of the absorption coefficient at different photon wavelengths with a uniaxial strain applied in the (a) [110] and (b) [111] directions.}
\label{fig:Absorption_strain}
\end{figure}

In addition to the shift of the cutoff wavelength, one can see that the absorption coefficient increases by a factor of 15 to 55 at the relaxed cutoff wavelength of 1.14~$\upmu$m, as better shown in Fig.~\ref{fig:Absorption_strain} showing the variation of the absorption coefficient with the applied strain, for different wavelengths.
While the absorption coefficient does not show large variations for wavelengths below 1~$\upmu$m, the enhancement is important near the cutoff wavelength (1.14~$\upmu$m).
For instance, the absorption coefficient increases by a factor of 55 when a $-2$\% strain is applied in the [110] direction.
The efficiency of a device detecting infrared photons would therefore be greatly enhanced, demonstrating the interest for highly strained silicon for infrared applications. 

\begin{figure}
    \includegraphics[width=1\linewidth]{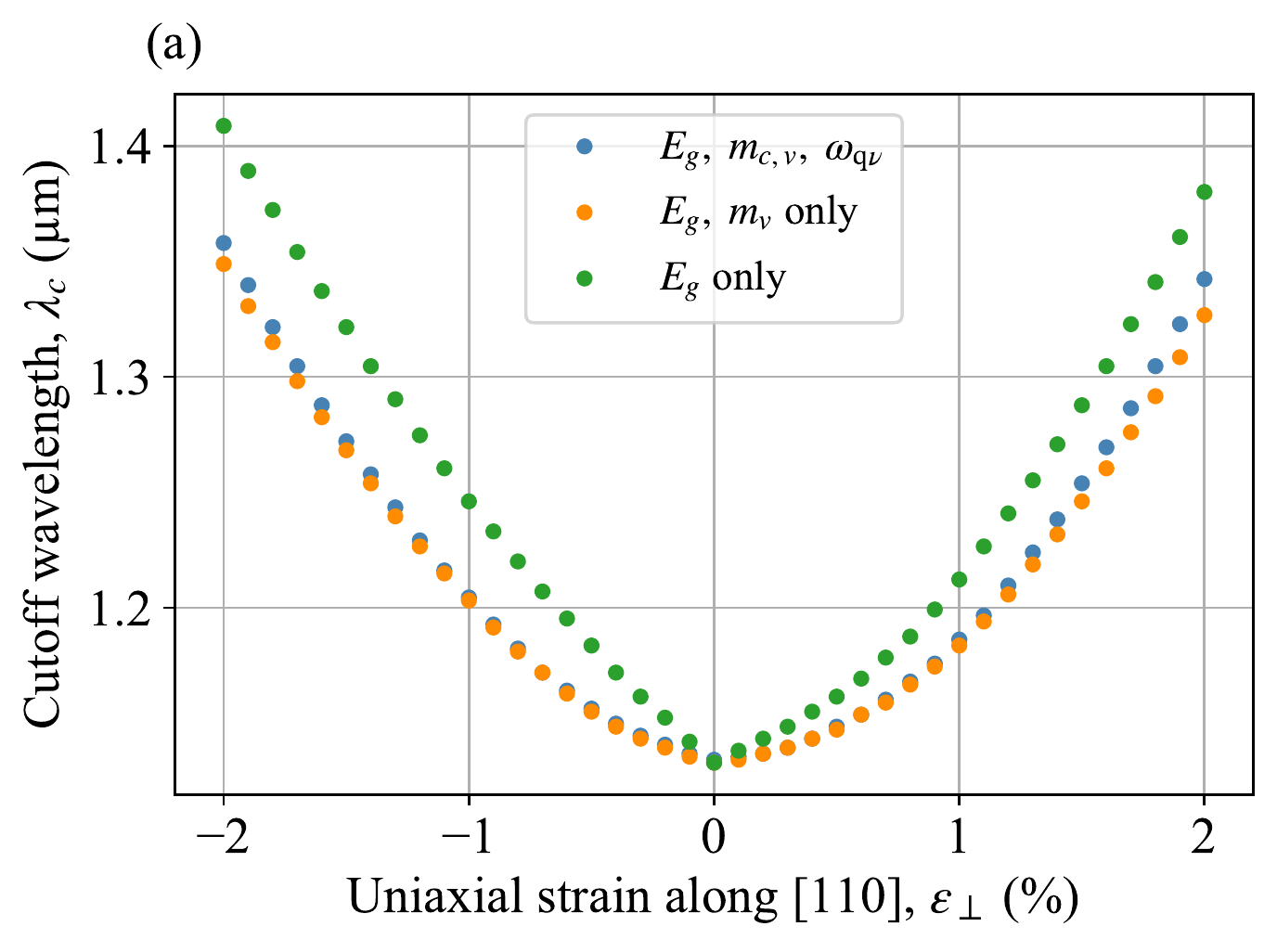}
    \includegraphics[width=1\linewidth]{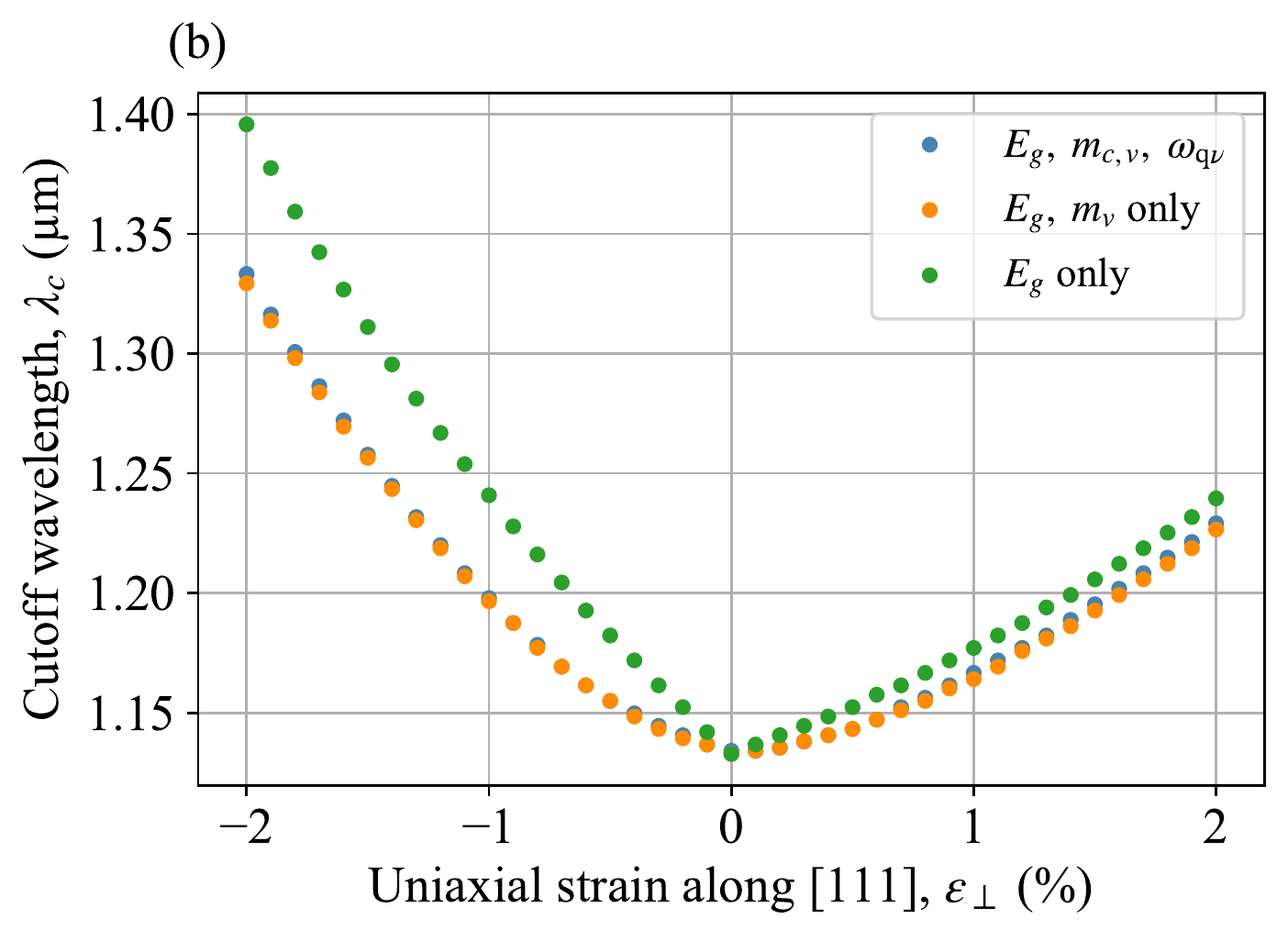}
    \caption{Evolution of the cutoff wavelength with a uniaxial strain applied in the (a) [110] and (b) [111] directions. Only the band gap changes with the strain are included (green), the valence band effective mass changes are also included (orange), or both effective masses and the phonon frequencies changes are also included (blue).}
\label{fig:CutOff}
\end{figure}

Another important impact of strain, besides the increased absorption in the infrared, is the shift of the cutoff wavelength towards higher wavelengths, as shown more precisely in Fig.~\ref{fig:CutOff}.
The changes due to the most important parameters are shown in this figure. 
Considering only the variation of the band gaps (in green) leads to an overestimation of the cutoff wavelength. 
A result very close to the one calculated with all the parameters changing (in blue) can be obtained by considering the variation of the valence band effective mass in addition to the band gaps (in orange). 
The impact of the changes in the conduction band effective masses and in phonon frequencies are negligible for most practical applications. 
We infer that, for a given strain tensor, it is sufficient to compute the relevant band gaps and valence band effective mass of strained silicon to determine its absorption coefficient at a very affordable computational cost.

\section{Conclusion}
\label{sec:conclusion}

In this work, we have adapted the model developed by Tsai\cite{Tsai2018} to compute the absorption coefficient of highly strained silicon.
Considering uniaxial strain along two directions, [110] and [111], we have computed the relevant material parameters for different strain levels up to $\pm2$\%.
This has allowed us to identify the band gap as the most important parameter changing with strain, directly followed by the valence band density-of-states effective mass. In contrast, the conduction band density-of-states effective masses and the phonon energies play a minor role in the absorption change.
We have shown that highly strained silicon can be used to extend the material bandwidth from 1.14 to 1.35~$\upmu$m.
Moreover, the absorption coefficient increases by a factor of 55 at the relaxed cutoff wavelength under a $-2$\% strain in the [110] direction. We have shown that the impact of strain is mainly significant around the cutoff wavelength which makes strain engineering particularly interesting for infrared applications.
This model can be used in conjunction with DFT computations of the relevant band gaps and of the valence band effective mass to determine the absorption coefficient of highly strained silicon in any configuration, at a very limited computational cost compared with fully first-principles electron-phonon computations.

\section*{Acknowledgments}

G.-M.R.~and G.B.~are grateful to the F.R.S.-FNRS (Belgium) and to the Communauté française de Belgique through the SURFASCOPE project (ARC 19/24-102) for financial support.

Computational resources have been provided by the supercomputing facilities of the Université catholique de Louvain (CISM/UCL) and the Consortium des Équipements de Calcul Intensif en Fédération Wallonie Bruxelles (CÉCI) funded by the Fond de la Recherche Scientifique de Belgique (F.R.S.-FNRS) under convention 2.5020.11 and by the Walloon Region.
\section*{Credit Lines}

This article may be downloaded for personal use only. Any other use requires prior permission of the author and AIP Publishing. This article may be found at \url{https://aip.scitation.org/doi/10.1063/5.0057350}.

\section*{Data Availability}

The data that supports the findings of this study are available within the article.

\bibliography{biblio_v1}

\begin{thebibliography}{51}%
\makeatletter
\providecommand \@ifxundefined [1]{%
 \@ifx{#1\undefined}
}%
\providecommand \@ifnum [1]{%
 \ifnum #1\expandafter \@firstoftwo
 \else \expandafter \@secondoftwo
 \fi
}%
\providecommand \@ifx [1]{%
 \ifx #1\expandafter \@firstoftwo
 \else \expandafter \@secondoftwo
 \fi
}%
\providecommand \natexlab [1]{#1}%
\providecommand \enquote  [1]{``#1''}%
\providecommand \bibnamefont  [1]{#1}%
\providecommand \bibfnamefont [1]{#1}%
\providecommand \citenamefont [1]{#1}%
\providecommand \href@noop [0]{\@secondoftwo}%
\providecommand \href [0]{\begingroup \@sanitize@url \@href}%
\providecommand \@href[1]{\@@startlink{#1}\@@href}%
\providecommand \@@href[1]{\endgroup#1\@@endlink}%
\providecommand \@sanitize@url [0]{\catcode `\\12\catcode `\$12\catcode
  `\&12\catcode `\#12\catcode `\^12\catcode `\_12\catcode `\%12\relax}%
\providecommand \@@startlink[1]{}%
\providecommand \@@endlink[0]{}%
\providecommand \url  [0]{\begingroup\@sanitize@url \@url }%
\providecommand \@url [1]{\endgroup\@href {#1}{\urlprefix }}%
\providecommand \urlprefix  [0]{URL }%
\providecommand \Eprint [0]{\href }%
\providecommand \doibase [0]{https://doi.org/}%
\providecommand \selectlanguage [0]{\@gobble}%
\providecommand \bibinfo  [0]{\@secondoftwo}%
\providecommand \bibfield  [0]{\@secondoftwo}%
\providecommand \translation [1]{[#1]}%
\providecommand \BibitemOpen [0]{}%
\providecommand \bibitemStop [0]{}%
\providecommand \bibitemNoStop [0]{.\EOS\space}%
\providecommand \EOS [0]{\spacefactor3000\relax}%
\providecommand \BibitemShut  [1]{\csname bibitem#1\endcsname}%
\let\auto@bib@innerbib\@empty
\bibitem [{\citenamefont {Licht}, \citenamefont {Peiró},\ and\ \citenamefont
  {Villalba}(2015)}]{Licht2015}%
  \BibitemOpen
  \bibfield  {author} {\bibinfo {author} {\bibfnamefont {C.}~\bibnamefont
  {Licht}}, \bibinfo {author} {\bibfnamefont {L.~T.}\ \bibnamefont {Peiró}},\
  and\ \bibinfo {author} {\bibfnamefont {G.}~\bibnamefont {Villalba}},\
  }\bibfield  {title} {\enquote {\bibinfo {title} {Global substance flow
  analysis of gallium, germanium, and indium: Quantification of extraction,
  uses, and dissipative losses within their anthropogenic cycles},}\ }\href
  {https://doi.org/https://doi.org/10.1111/jiec.12287} {\bibfield  {journal}
  {\bibinfo  {journal} {Journal of Industrial Ecology}\ }\textbf {\bibinfo
  {volume} {19}},\ \bibinfo {pages} {890--903} (\bibinfo {year} {2015})},\
  \Eprint
  {https://arxiv.org/abs/https://onlinelibrary.wiley.com/doi/pdf/10.1111/jiec.12287}
  {https://onlinelibrary.wiley.com/doi/pdf/10.1111/jiec.12287} \BibitemShut
  {NoStop}%
\bibitem [{\citenamefont {Warnock}(2011)}]{Warnock2011}%
  \BibitemOpen
  \bibfield  {author} {\bibinfo {author} {\bibfnamefont {J.}~\bibnamefont
  {Warnock}},\ }\bibfield  {title} {\enquote {\bibinfo {title} {Circuit design
  challenges at the 14nm technology node},}\ }in\ \href
  {https://doi.org/https://doi.org/10.1145/2024724.2024833} {\emph {\bibinfo
  {booktitle} {Proceedings of the 48th Design Automation Conference}}}\
  (\bibinfo {year} {2011})\ pp.\ \bibinfo {pages} {464--467}\BibitemShut
  {NoStop}%
\bibitem [{\citenamefont {Waldrop}(2016)}]{Waldrop2016}%
  \BibitemOpen
  \bibfield  {author} {\bibinfo {author} {\bibfnamefont {M.~M.}\ \bibnamefont
  {Waldrop}},\ }\bibfield  {title} {\enquote {\bibinfo {title} {More than
  moore},}\ }\href {https://doi.org/https://doi.org/10.1038/530144a} {\bibfield
   {journal} {\bibinfo  {journal} {Nature}\ }\textbf {\bibinfo {volume}
  {530}},\ \bibinfo {pages} {144--148} (\bibinfo {year} {2016})}\BibitemShut
  {NoStop}%
\bibitem [{\citenamefont {Jaeger}\ and\ \citenamefont
  {Suhling}(2018)}]{Jaeger2018}%
  \BibitemOpen
  \bibfield  {author} {\bibinfo {author} {\bibfnamefont {R.~C.}\ \bibnamefont
  {Jaeger}}\ and\ \bibinfo {author} {\bibfnamefont {J.~C.}\ \bibnamefont
  {Suhling}},\ }\bibfield  {title} {\enquote {\bibinfo {title} {First and
  second order piezoresistive characteristics of cmos fets: Weak through strong
  inversion},}\ }in\ \href {https://doi.org/10.1109/ESSDERC.2018.8486881}
  {\emph {\bibinfo {booktitle} {2018 48th European Solid-State Device Research
  Conference (ESSDERC)}}}\ (\bibinfo {organization} {IEEE},\ \bibinfo {year}
  {2018})\ pp.\ \bibinfo {pages} {126--129}\BibitemShut {NoStop}%
\bibitem [{\citenamefont {Heremans}\ \emph {et~al.}(2016)\citenamefont
  {Heremans}, \citenamefont {Tripathi}, \citenamefont {de~Jamblinne~de Meux},
  \citenamefont {Smits}, \citenamefont {Hou}, \citenamefont {Pourtois},\ and\
  \citenamefont {Gelinck}}]{Heremans2016}%
  \BibitemOpen
  \bibfield  {author} {\bibinfo {author} {\bibfnamefont {P.}~\bibnamefont
  {Heremans}}, \bibinfo {author} {\bibfnamefont {A.~K.}\ \bibnamefont
  {Tripathi}}, \bibinfo {author} {\bibfnamefont {A.}~\bibnamefont
  {de~Jamblinne~de Meux}}, \bibinfo {author} {\bibfnamefont {E.~C.}\
  \bibnamefont {Smits}}, \bibinfo {author} {\bibfnamefont {B.}~\bibnamefont
  {Hou}}, \bibinfo {author} {\bibfnamefont {G.}~\bibnamefont {Pourtois}},\ and\
  \bibinfo {author} {\bibfnamefont {G.~H.}\ \bibnamefont {Gelinck}},\
  }\bibfield  {title} {\enquote {\bibinfo {title} {Mechanical and electronic
  properties of thin-film transistors on plastic, and their integration in
  flexible electronic applications},}\ }\href
  {https://doi.org/https://doi.org/10.1002/adma.201504360} {\bibfield
  {journal} {\bibinfo  {journal} {Advanced Materials}\ }\textbf {\bibinfo
  {volume} {28}},\ \bibinfo {pages} {4266--4282} (\bibinfo {year}
  {2016})}\BibitemShut {NoStop}%
\bibitem [{\citenamefont {Reiche}\ \emph {et~al.}(2009)\citenamefont {Reiche},
  \citenamefont {Moutanabbir}, \citenamefont {Hoentschel}, \citenamefont
  {G{\"{o}}sele}, \citenamefont {Flachowsky},\ and\ \citenamefont
  {Horstmann}}]{Reiche2009}%
  \BibitemOpen
  \bibfield  {author} {\bibinfo {author} {\bibfnamefont {M.}~\bibnamefont
  {Reiche}}, \bibinfo {author} {\bibfnamefont {O.}~\bibnamefont {Moutanabbir}},
  \bibinfo {author} {\bibfnamefont {J.}~\bibnamefont {Hoentschel}}, \bibinfo
  {author} {\bibfnamefont {U.}~\bibnamefont {G{\"{o}}sele}}, \bibinfo {author}
  {\bibfnamefont {S.}~\bibnamefont {Flachowsky}},\ and\ \bibinfo {author}
  {\bibfnamefont {M.}~\bibnamefont {Horstmann}},\ }\bibfield  {title} {\enquote
  {\bibinfo {title} {{Strained Silicon Devices}},}\ }\href
  {https://doi.org/10.4028/www.scientific.net/SSP.156-158.61} {\bibfield
  {journal} {\bibinfo  {journal} {Solid State Phenomena}\ }\textbf {\bibinfo
  {volume} {156-158}},\ \bibinfo {pages} {61--68} (\bibinfo {year}
  {2009})}\BibitemShut {NoStop}%
\bibitem [{\citenamefont {Gnanachchelvi}\ \emph {et~al.}(2016)\citenamefont
  {Gnanachchelvi}, \citenamefont {Jaeger}, \citenamefont {Wilamowski},
  \citenamefont {Niu}, \citenamefont {Hussain}, \citenamefont {Suhling},\ and\
  \citenamefont {Hamilton}}]{Gnanachchelvi2016}%
  \BibitemOpen
  \bibfield  {author} {\bibinfo {author} {\bibfnamefont {P.}~\bibnamefont
  {Gnanachchelvi}}, \bibinfo {author} {\bibfnamefont {R.}~\bibnamefont
  {Jaeger}}, \bibinfo {author} {\bibfnamefont {B.}~\bibnamefont {Wilamowski}},
  \bibinfo {author} {\bibfnamefont {G.}~\bibnamefont {Niu}}, \bibinfo {author}
  {\bibfnamefont {S.}~\bibnamefont {Hussain}}, \bibinfo {author} {\bibfnamefont
  {J.}~\bibnamefont {Suhling}},\ and\ \bibinfo {author} {\bibfnamefont
  {M.}~\bibnamefont {Hamilton}},\ }\bibfield  {title} {\enquote {\bibinfo
  {title} {Performance enhancement in bipolar junction transistors using
  uniaxial stress on (100) silicon},}\ }\href
  {https://doi.org/10.1109/TED.2016.2560899} {\bibfield  {journal} {\bibinfo
  {journal} {IEEE Transactions on Electron Devices}\ }\textbf {\bibinfo
  {volume} {63}},\ \bibinfo {pages} {2643--2649} (\bibinfo {year}
  {2016})}\BibitemShut {NoStop}%
\bibitem [{\citenamefont {Thompson}\ \emph {et~al.}(2006)\citenamefont
  {Thompson}, \citenamefont {Suthram}, \citenamefont {Sun}, \citenamefont
  {Sun}, \citenamefont {Parthasarathy}, \citenamefont {Chu},\ and\
  \citenamefont {Nishida}}]{Thompson2006}%
  \BibitemOpen
  \bibfield  {author} {\bibinfo {author} {\bibfnamefont {S.}~\bibnamefont
  {Thompson}}, \bibinfo {author} {\bibfnamefont {S.}~\bibnamefont {Suthram}},
  \bibinfo {author} {\bibfnamefont {Y.}~\bibnamefont {Sun}}, \bibinfo {author}
  {\bibfnamefont {G.}~\bibnamefont {Sun}}, \bibinfo {author} {\bibfnamefont
  {S.}~\bibnamefont {Parthasarathy}}, \bibinfo {author} {\bibfnamefont
  {M.}~\bibnamefont {Chu}},\ and\ \bibinfo {author} {\bibfnamefont
  {T.}~\bibnamefont {Nishida}},\ }\bibfield  {title} {\enquote {\bibinfo
  {title} {Future of strained si/semiconductors in nanoscale mosfets},}\ }in\
  \href {https://doi.org/10.1109/IEDM.2006.346877} {\emph {\bibinfo {booktitle}
  {2006 International Electron Devices Meeting}}}\ (\bibinfo {organization}
  {IEEE},\ \bibinfo {year} {2006})\ pp.\ \bibinfo {pages} {1--4}\BibitemShut
  {NoStop}%
\bibitem [{\citenamefont {Kleimann}\ \emph {et~al.}(1998)\citenamefont
  {Kleimann}, \citenamefont {Semmache}, \citenamefont {Le~Berre},\ and\
  \citenamefont {Barbier}}]{Kleimann1998}%
  \BibitemOpen
  \bibfield  {author} {\bibinfo {author} {\bibfnamefont {P.}~\bibnamefont
  {Kleimann}}, \bibinfo {author} {\bibfnamefont {B.}~\bibnamefont {Semmache}},
  \bibinfo {author} {\bibfnamefont {M.}~\bibnamefont {Le~Berre}},\ and\
  \bibinfo {author} {\bibfnamefont {D.}~\bibnamefont {Barbier}},\ }\bibfield
  {title} {\enquote {\bibinfo {title} {Stress-dependent hole effective masses
  and piezoresistive properties of p-type monocrystalline and polycrystalline
  silicon},}\ }\href {https://doi.org/https://doi.org/10.1103/PhysRevB.57.8966}
  {\bibfield  {journal} {\bibinfo  {journal} {Physical Review B}\ }\textbf
  {\bibinfo {volume} {57}},\ \bibinfo {pages} {8966} (\bibinfo {year}
  {1998})}\BibitemShut {NoStop}%
\bibitem [{\citenamefont {Nielsen}\ and\ \citenamefont
  {Martin}(1985)}]{Nielsen1985}%
  \BibitemOpen
  \bibfield  {author} {\bibinfo {author} {\bibfnamefont {O.}~\bibnamefont
  {Nielsen}}\ and\ \bibinfo {author} {\bibfnamefont {R.~M.}\ \bibnamefont
  {Martin}},\ }\bibfield  {title} {\enquote {\bibinfo {title} {Stresses in
  semiconductors: Ab initio calculations on si, ge, and gaas},}\ }\href
  {https://doi.org/https://doi.org/10.1103/PhysRevB.32.3792} {\bibfield
  {journal} {\bibinfo  {journal} {Physical Review B}\ }\textbf {\bibinfo
  {volume} {32}},\ \bibinfo {pages} {3792} (\bibinfo {year}
  {1985})}\BibitemShut {NoStop}%
\bibitem [{\citenamefont {Rieger}\ and\ \citenamefont
  {Vogl}(1993)}]{Rieger1993}%
  \BibitemOpen
  \bibfield  {author} {\bibinfo {author} {\bibfnamefont {M.~M.}\ \bibnamefont
  {Rieger}}\ and\ \bibinfo {author} {\bibfnamefont {P.}~\bibnamefont {Vogl}},\
  }\bibfield  {title} {\enquote {\bibinfo {title} {Electronic-band parameters
  in strained si 1- x ge x alloys on si 1- y ge y substrates},}\ }\href
  {https://doi.org/https://doi.org/10.1103/PhysRevB.48.14276} {\bibfield
  {journal} {\bibinfo  {journal} {Physical Review B}\ }\textbf {\bibinfo
  {volume} {48}},\ \bibinfo {pages} {14276} (\bibinfo {year}
  {1993})}\BibitemShut {NoStop}%
\bibitem [{\citenamefont {Goroff}\ and\ \citenamefont
  {Kleinman}(1963)}]{Goroff1963}%
  \BibitemOpen
  \bibfield  {author} {\bibinfo {author} {\bibfnamefont {I.}~\bibnamefont
  {Goroff}}\ and\ \bibinfo {author} {\bibfnamefont {L.}~\bibnamefont
  {Kleinman}},\ }\bibfield  {title} {\enquote {\bibinfo {title} {Deformation
  potentials in silicon. iii. effects of a general strain on conduction and
  valence levels},}\ }\href
  {https://doi.org/https://doi.org/10.1103/PhysRev.132.1080} {\bibfield
  {journal} {\bibinfo  {journal} {Physical Review}\ }\textbf {\bibinfo {volume}
  {132}},\ \bibinfo {pages} {1080} (\bibinfo {year} {1963})}\BibitemShut
  {NoStop}%
\bibitem [{\citenamefont {Passi}\ \emph {et~al.}(2012)\citenamefont {Passi},
  \citenamefont {Bhaskar}, \citenamefont {Pardoen}, \citenamefont {Sodervall},
  \citenamefont {Nilsson}, \citenamefont {Petersson}, \citenamefont {Hagberg},\
  and\ \citenamefont {Raskin}}]{Passi2012}%
  \BibitemOpen
  \bibfield  {author} {\bibinfo {author} {\bibfnamefont {V.}~\bibnamefont
  {Passi}}, \bibinfo {author} {\bibfnamefont {U.}~\bibnamefont {Bhaskar}},
  \bibinfo {author} {\bibfnamefont {T.}~\bibnamefont {Pardoen}}, \bibinfo
  {author} {\bibfnamefont {U.}~\bibnamefont {Sodervall}}, \bibinfo {author}
  {\bibfnamefont {B.}~\bibnamefont {Nilsson}}, \bibinfo {author} {\bibfnamefont
  {G.}~\bibnamefont {Petersson}}, \bibinfo {author} {\bibfnamefont
  {M.}~\bibnamefont {Hagberg}},\ and\ \bibinfo {author} {\bibfnamefont {J.-P.}\
  \bibnamefont {Raskin}},\ }\bibfield  {title} {\enquote {\bibinfo {title}
  {High-throughput on-chip large deformation of silicon nanoribbons and
  nanowires},}\ }\href {https://doi.org/10.1109/JMEMS.2012.2190711} {\bibfield
  {journal} {\bibinfo  {journal} {Journal of Microelectromechanical Systems}\
  }\textbf {\bibinfo {volume} {21}},\ \bibinfo {pages} {822--829} (\bibinfo
  {year} {2012})}\BibitemShut {NoStop}%
\bibitem [{\citenamefont {Bhaskar}\ \emph {et~al.}(2012)\citenamefont
  {Bhaskar}, \citenamefont {Passi}, \citenamefont {Houri}, \citenamefont
  {Escobedo-Cousin}, \citenamefont {Olsen}, \citenamefont {Pardoen},\ and\
  \citenamefont {Raskin}}]{Bhaskar2012}%
  \BibitemOpen
  \bibfield  {author} {\bibinfo {author} {\bibfnamefont {U.}~\bibnamefont
  {Bhaskar}}, \bibinfo {author} {\bibfnamefont {V.}~\bibnamefont {Passi}},
  \bibinfo {author} {\bibfnamefont {S.}~\bibnamefont {Houri}}, \bibinfo
  {author} {\bibfnamefont {E.}~\bibnamefont {Escobedo-Cousin}}, \bibinfo
  {author} {\bibfnamefont {S.~H.}\ \bibnamefont {Olsen}}, \bibinfo {author}
  {\bibfnamefont {T.}~\bibnamefont {Pardoen}},\ and\ \bibinfo {author}
  {\bibfnamefont {J.-P.}\ \bibnamefont {Raskin}},\ }\bibfield  {title}
  {\enquote {\bibinfo {title} {On-chip tensile testing of nanoscale silicon
  free-standing beams},}\ }\href
  {https://doi.org/https://doi.org/10.1557/jmr.2011.340} {\bibfield  {journal}
  {\bibinfo  {journal} {Journal of Materials Research}\ }\textbf {\bibinfo
  {volume} {27}},\ \bibinfo {pages} {571--579} (\bibinfo {year}
  {2012})}\BibitemShut {NoStop}%
\bibitem [{\citenamefont {{Kumar Bhaskar}}\ \emph {et~al.}(2013)\citenamefont
  {{Kumar Bhaskar}}, \citenamefont {Pardoen}, \citenamefont {Passi},\ and\
  \citenamefont {Raskin}}]{KumarBhaskar2013}%
  \BibitemOpen
  \bibfield  {author} {\bibinfo {author} {\bibfnamefont {U.}~\bibnamefont
  {{Kumar Bhaskar}}}, \bibinfo {author} {\bibfnamefont {T.}~\bibnamefont
  {Pardoen}}, \bibinfo {author} {\bibfnamefont {V.}~\bibnamefont {Passi}},\
  and\ \bibinfo {author} {\bibfnamefont {J.-P.}\ \bibnamefont {Raskin}},\
  }\bibfield  {title} {\enquote {\bibinfo {title} {{Piezoresistance of
  nano-scale silicon up to 2 GPa in tension}},}\ }\href
  {https://doi.org/10.1063/1.4788919} {\bibfield  {journal} {\bibinfo
  {journal} {Applied Physics Letters}\ }\textbf {\bibinfo {volume} {102}},\
  \bibinfo {pages} {031911} (\bibinfo {year} {2013})}\BibitemShut {NoStop}%
\bibitem [{\citenamefont {Zhang}\ \emph {et~al.}(2016)\citenamefont {Zhang},
  \citenamefont {Tersoff}, \citenamefont {Xu}, \citenamefont {Chen},
  \citenamefont {Zhang}, \citenamefont {Zhang}, \citenamefont {Yang},
  \citenamefont {Lee}, \citenamefont {Tu}, \citenamefont {Li} \emph
  {et~al.}}]{Zhang2016}%
  \BibitemOpen
  \bibfield  {author} {\bibinfo {author} {\bibfnamefont {H.}~\bibnamefont
  {Zhang}}, \bibinfo {author} {\bibfnamefont {J.}~\bibnamefont {Tersoff}},
  \bibinfo {author} {\bibfnamefont {S.}~\bibnamefont {Xu}}, \bibinfo {author}
  {\bibfnamefont {H.}~\bibnamefont {Chen}}, \bibinfo {author} {\bibfnamefont
  {Q.}~\bibnamefont {Zhang}}, \bibinfo {author} {\bibfnamefont
  {K.}~\bibnamefont {Zhang}}, \bibinfo {author} {\bibfnamefont
  {Y.}~\bibnamefont {Yang}}, \bibinfo {author} {\bibfnamefont {C.-S.}\
  \bibnamefont {Lee}}, \bibinfo {author} {\bibfnamefont {K.-N.}\ \bibnamefont
  {Tu}}, \bibinfo {author} {\bibfnamefont {J.}~\bibnamefont {Li}}, \emph
  {et~al.},\ }\bibfield  {title} {\enquote {\bibinfo {title} {Approaching the
  ideal elastic strain limit in silicon nanowires},}\ }\href
  {https://doi.org/10.1126/sciadv.1501382} {\bibfield  {journal} {\bibinfo
  {journal} {Science Advances}\ }\textbf {\bibinfo {volume} {2}},\ \bibinfo
  {pages} {e1501382} (\bibinfo {year} {2016})}\BibitemShut {NoStop}%
\bibitem [{\citenamefont {Cavallo}\ and\ \citenamefont
  {Lagally}(2012)}]{Cavallo2012}%
  \BibitemOpen
  \bibfield  {author} {\bibinfo {author} {\bibfnamefont {F.}~\bibnamefont
  {Cavallo}}\ and\ \bibinfo {author} {\bibfnamefont {M.~G.}\ \bibnamefont
  {Lagally}},\ }\bibfield  {title} {\enquote {\bibinfo {title} {{Semiconductor
  nanomembranes: A platform for new properties via strain engineering}},}\
  }\href {https://doi.org/10.1186/1556-276X-7-628} {\bibfield  {journal}
  {\bibinfo  {journal} {Nanoscale Research Letters}\ }\textbf {\bibinfo
  {volume} {7}},\ \bibinfo {pages} {1--10} (\bibinfo {year}
  {2012})}\BibitemShut {NoStop}%
\bibitem [{\citenamefont {Montmeat}\ \emph {et~al.}(2016)\citenamefont
  {Montmeat}, \citenamefont {{De Nigris Brandolisi}}, \citenamefont {Tardif},
  \citenamefont {Enot}, \citenamefont {Enyedi}, \citenamefont {Kachtouli},
  \citenamefont {Besson}, \citenamefont {Rieutord},\ and\ \citenamefont
  {Fournel}}]{Montmeat2016}%
  \BibitemOpen
  \bibfield  {author} {\bibinfo {author} {\bibfnamefont {P.}~\bibnamefont
  {Montmeat}}, \bibinfo {author} {\bibfnamefont {I.}~\bibnamefont {{De Nigris
  Brandolisi}}}, \bibinfo {author} {\bibfnamefont {S.}~\bibnamefont {Tardif}},
  \bibinfo {author} {\bibfnamefont {T.}~\bibnamefont {Enot}}, \bibinfo {author}
  {\bibfnamefont {G.}~\bibnamefont {Enyedi}}, \bibinfo {author} {\bibfnamefont
  {R.}~\bibnamefont {Kachtouli}}, \bibinfo {author} {\bibfnamefont
  {P.}~\bibnamefont {Besson}}, \bibinfo {author} {\bibfnamefont
  {F.}~\bibnamefont {Rieutord}},\ and\ \bibinfo {author} {\bibfnamefont
  {F.}~\bibnamefont {Fournel}},\ }\bibfield  {title} {\enquote {\bibinfo
  {title} {{Transfer of Ultra-Thin Semi-Conductor Films onto Flexible
  Substrates}},}\ }\href {https://doi.org/10.1149/07509.0247ecst} {\bibfield
  {journal} {\bibinfo  {journal} {ECS Transactions}\ }\textbf {\bibinfo
  {volume} {75}},\ \bibinfo {pages} {247--252} (\bibinfo {year}
  {2016})}\BibitemShut {NoStop}%
\bibitem [{\citenamefont {Mungu{\'\i}a}\ \emph {et~al.}(2008)\citenamefont
  {Mungu{\'\i}a}, \citenamefont {Bremond}, \citenamefont {Bluet}, \citenamefont
  {Hartmann},\ and\ \citenamefont {Mermoux}}]{Munguia2008}%
  \BibitemOpen
  \bibfield  {author} {\bibinfo {author} {\bibfnamefont {J.}~\bibnamefont
  {Mungu{\'\i}a}}, \bibinfo {author} {\bibfnamefont {G.}~\bibnamefont
  {Bremond}}, \bibinfo {author} {\bibfnamefont {J.}~\bibnamefont {Bluet}},
  \bibinfo {author} {\bibfnamefont {J.}~\bibnamefont {Hartmann}},\ and\
  \bibinfo {author} {\bibfnamefont {M.}~\bibnamefont {Mermoux}},\ }\bibfield
  {title} {\enquote {\bibinfo {title} {Strain dependence of indirect band gap
  for strained silicon on insulator wafers},}\ }\href
  {https://doi.org/https://doi.org/10.1063/1.2978241} {\bibfield  {journal}
  {\bibinfo  {journal} {Applied Physics Letters}\ }\textbf {\bibinfo {volume}
  {93}},\ \bibinfo {pages} {102101} (\bibinfo {year} {2008})}\BibitemShut
  {NoStop}%
\bibitem [{\citenamefont {Lange}\ \emph {et~al.}(2016)\citenamefont {Lange},
  \citenamefont {i~Cabarrocas}, \citenamefont {Triantafyllidis},\ and\
  \citenamefont {Daineka}}]{Lange2016}%
  \BibitemOpen
  \bibfield  {author} {\bibinfo {author} {\bibfnamefont {D.}~\bibnamefont
  {Lange}}, \bibinfo {author} {\bibfnamefont {P.~R.}\ \bibnamefont
  {i~Cabarrocas}}, \bibinfo {author} {\bibfnamefont {N.}~\bibnamefont
  {Triantafyllidis}},\ and\ \bibinfo {author} {\bibfnamefont {D.}~\bibnamefont
  {Daineka}},\ }\bibfield  {title} {\enquote {\bibinfo {title}
  {Piezoresistivity of thin film semiconductors with application to thin film
  silicon solar cells},}\ }\href
  {https://doi.org/https://doi.org/10.1016/j.solmat.2015.09.014} {\bibfield
  {journal} {\bibinfo  {journal} {Solar Energy Materials and Solar Cells}\
  }\textbf {\bibinfo {volume} {145}},\ \bibinfo {pages} {93--103} (\bibinfo
  {year} {2016})}\BibitemShut {NoStop}%
\bibitem [{\citenamefont {Schriever}\ and\ \citenamefont
  {Wehrspohn}(2012)}]{Schriever2012}%
  \BibitemOpen
  \bibfield  {author} {\bibinfo {author} {\bibfnamefont {C.}~\bibnamefont
  {Schriever}}\ and\ \bibinfo {author} {\bibfnamefont {R.~B.}\ \bibnamefont
  {Wehrspohn}},\ }\bibfield  {title} {\enquote {\bibinfo {title} {Stretching
  silicon's potential},}\ }\href
  {https://doi.org/https://doi.org/10.1038/nmat3226} {\bibfield  {journal}
  {\bibinfo  {journal} {Nature Materials}\ }\textbf {\bibinfo {volume} {11}},\
  \bibinfo {pages} {96--97} (\bibinfo {year} {2012})}\BibitemShut {NoStop}%
\bibitem [{\citenamefont {Cazzanelli}\ \emph {et~al.}(2012)\citenamefont
  {Cazzanelli}, \citenamefont {Bianco}, \citenamefont {Borga}, \citenamefont
  {Pucker}, \citenamefont {Ghulinyan}, \citenamefont {Degoli}, \citenamefont
  {Luppi}, \citenamefont {V{\'e}niard}, \citenamefont {Ossicini}, \citenamefont
  {Modotto} \emph {et~al.}}]{Cazzanelli2012}%
  \BibitemOpen
  \bibfield  {author} {\bibinfo {author} {\bibfnamefont {M.}~\bibnamefont
  {Cazzanelli}}, \bibinfo {author} {\bibfnamefont {F.}~\bibnamefont {Bianco}},
  \bibinfo {author} {\bibfnamefont {E.}~\bibnamefont {Borga}}, \bibinfo
  {author} {\bibfnamefont {G.}~\bibnamefont {Pucker}}, \bibinfo {author}
  {\bibfnamefont {M.}~\bibnamefont {Ghulinyan}}, \bibinfo {author}
  {\bibfnamefont {E.}~\bibnamefont {Degoli}}, \bibinfo {author} {\bibfnamefont
  {E.}~\bibnamefont {Luppi}}, \bibinfo {author} {\bibfnamefont
  {V.}~\bibnamefont {V{\'e}niard}}, \bibinfo {author} {\bibfnamefont
  {S.}~\bibnamefont {Ossicini}}, \bibinfo {author} {\bibfnamefont
  {D.}~\bibnamefont {Modotto}}, \emph {et~al.},\ }\bibfield  {title} {\enquote
  {\bibinfo {title} {Second-harmonic generation in silicon waveguides strained
  by silicon nitride},}\ }\href
  {https://doi.org/https://doi.org/10.1038/nmat3200} {\bibfield  {journal}
  {\bibinfo  {journal} {Nature Materials}\ }\textbf {\bibinfo {volume} {11}},\
  \bibinfo {pages} {148--154} (\bibinfo {year} {2012})}\BibitemShut {NoStop}%
\bibitem [{\citenamefont {Fischetti}\ and\ \citenamefont
  {Laux}(1996)}]{Fischetti1996}%
  \BibitemOpen
  \bibfield  {author} {\bibinfo {author} {\bibfnamefont {M.~V.}\ \bibnamefont
  {Fischetti}}\ and\ \bibinfo {author} {\bibfnamefont {S.~E.}\ \bibnamefont
  {Laux}},\ }\bibfield  {title} {\enquote {\bibinfo {title} {Band structure,
  deformation potentials, and carrier mobility in strained si, ge, and sige
  alloys},}\ }\href {https://doi.org/https://doi.org/10.1063/1.363052}
  {\bibfield  {journal} {\bibinfo  {journal} {Journal of Applied Physics}\
  }\textbf {\bibinfo {volume} {80}},\ \bibinfo {pages} {2234--2252} (\bibinfo
  {year} {1996})}\BibitemShut {NoStop}%
\bibitem [{\citenamefont {Wen}\ and\ \citenamefont {Bellotti}(2015)}]{Wen2015}%
  \BibitemOpen
  \bibfield  {author} {\bibinfo {author} {\bibfnamefont {H.}~\bibnamefont
  {Wen}}\ and\ \bibinfo {author} {\bibfnamefont {E.}~\bibnamefont {Bellotti}},\
  }\bibfield  {title} {\enquote {\bibinfo {title} {Rigorous theory of the
  radiative and gain characteristics of silicon and germanium lasing media},}\
  }\href {https://doi.org/10.1103/PhysRevB.91.035307} {\bibfield  {journal}
  {\bibinfo  {journal} {Physical Review B}\ }\textbf {\bibinfo {volume} {91}},\
  \bibinfo {pages} {035307} (\bibinfo {year} {2015})}\BibitemShut {NoStop}%
\bibitem [{\citenamefont {Lin}\ \emph {et~al.}(2017)\citenamefont {Lin},
  \citenamefont {Luo}, \citenamefont {Gu}, \citenamefont {Kimerling},
  \citenamefont {Wada}, \citenamefont {Agarwal},\ and\ \citenamefont
  {Hu}}]{Lin2017}%
  \BibitemOpen
  \bibfield  {author} {\bibinfo {author} {\bibfnamefont {H.}~\bibnamefont
  {Lin}}, \bibinfo {author} {\bibfnamefont {Z.}~\bibnamefont {Luo}}, \bibinfo
  {author} {\bibfnamefont {T.}~\bibnamefont {Gu}}, \bibinfo {author}
  {\bibfnamefont {L.~C.}\ \bibnamefont {Kimerling}}, \bibinfo {author}
  {\bibfnamefont {K.}~\bibnamefont {Wada}}, \bibinfo {author} {\bibfnamefont
  {A.}~\bibnamefont {Agarwal}},\ and\ \bibinfo {author} {\bibfnamefont
  {J.}~\bibnamefont {Hu}},\ }\bibfield  {title} {\enquote {\bibinfo {title}
  {Mid-infrared integrated photonics on silicon: a perspective},}\ }\href
  {https://doi.org/https://doi.org/10.1515/nanoph-2017-0085} {\bibfield
  {journal} {\bibinfo  {journal} {Nanophotonics}\ }\textbf {\bibinfo {volume}
  {7}},\ \bibinfo {pages} {393--420} (\bibinfo {year} {2017})}\BibitemShut
  {NoStop}%
\bibitem [{\citenamefont {Noffsinger}\ \emph {et~al.}(2012)\citenamefont
  {Noffsinger}, \citenamefont {Kioupakis}, \citenamefont {Van~de Walle},
  \citenamefont {Louie},\ and\ \citenamefont {Cohen}}]{Noffsinger2012}%
  \BibitemOpen
  \bibfield  {author} {\bibinfo {author} {\bibfnamefont {J.}~\bibnamefont
  {Noffsinger}}, \bibinfo {author} {\bibfnamefont {E.}~\bibnamefont
  {Kioupakis}}, \bibinfo {author} {\bibfnamefont {C.~G.}\ \bibnamefont {Van~de
  Walle}}, \bibinfo {author} {\bibfnamefont {S.~G.}\ \bibnamefont {Louie}},\
  and\ \bibinfo {author} {\bibfnamefont {M.~L.}\ \bibnamefont {Cohen}},\
  }\bibfield  {title} {\enquote {\bibinfo {title} {Phonon-assisted optical
  absorption in silicon from first principles},}\ }\href
  {https://doi.org/https://doi.org/10.1103/PhysRevLett.108.167402} {\bibfield
  {journal} {\bibinfo  {journal} {Physical Review Letters}\ }\textbf {\bibinfo
  {volume} {108}},\ \bibinfo {pages} {167402} (\bibinfo {year}
  {2012})}\BibitemShut {NoStop}%
\bibitem [{\citenamefont {Zacharias}, \citenamefont {Patrick},\ and\
  \citenamefont {Giustino}(2015)}]{Zacharias2015}%
  \BibitemOpen
  \bibfield  {author} {\bibinfo {author} {\bibfnamefont {M.}~\bibnamefont
  {Zacharias}}, \bibinfo {author} {\bibfnamefont {C.~E.}\ \bibnamefont
  {Patrick}},\ and\ \bibinfo {author} {\bibfnamefont {F.}~\bibnamefont
  {Giustino}},\ }\bibfield  {title} {\enquote {\bibinfo {title} {Stochastic
  approach to phonon-assisted optical absorption},}\ }\href
  {https://doi.org/https://doi.org/10.1103/PhysRevLett.115.177401} {\bibfield
  {journal} {\bibinfo  {journal} {Physical Review Letters}\ }\textbf {\bibinfo
  {volume} {115}},\ \bibinfo {pages} {177401} (\bibinfo {year}
  {2015})}\BibitemShut {NoStop}%
\bibitem [{\citenamefont {Tsai}(2018)}]{Tsai2018}%
  \BibitemOpen
  \bibfield  {author} {\bibinfo {author} {\bibfnamefont {C.-Y.}\ \bibnamefont
  {Tsai}},\ }\bibfield  {title} {\enquote {\bibinfo {title} {Absorption
  coefficients of silicon: A theoretical treatment},}\ }\href
  {https://doi.org/https://doi.org/10.1063/1.5028053} {\bibfield  {journal}
  {\bibinfo  {journal} {Journal of Applied Physics}\ }\textbf {\bibinfo
  {volume} {123}},\ \bibinfo {pages} {183103} (\bibinfo {year}
  {2018})}\BibitemShut {NoStop}%
\bibitem [{\citenamefont {Patrick}\ and\ \citenamefont
  {Giustino}(2014)}]{Patrick2014}%
  \BibitemOpen
  \bibfield  {author} {\bibinfo {author} {\bibfnamefont {C.~E.}\ \bibnamefont
  {Patrick}}\ and\ \bibinfo {author} {\bibfnamefont {F.}~\bibnamefont
  {Giustino}},\ }\bibfield  {title} {\enquote {\bibinfo {title} {Unified theory
  of electron--phonon renormalization and phonon-assisted optical
  absorption},}\ }\href
  {https://doi.org/https://doi.org/10.1088/0953-8984/26/36/365503} {\bibfield
  {journal} {\bibinfo  {journal} {Journal of Physics: Condensed Matter}\
  }\textbf {\bibinfo {volume} {26}},\ \bibinfo {pages} {365503} (\bibinfo
  {year} {2014})}\BibitemShut {NoStop}%
\bibitem [{\citenamefont {Giustino}(2017)}]{Giustino2017}%
  \BibitemOpen
  \bibfield  {author} {\bibinfo {author} {\bibfnamefont {F.}~\bibnamefont
  {Giustino}},\ }\bibfield  {title} {\enquote {\bibinfo {title}
  {Electron-phonon interactions from first principles},}\ }\href
  {https://doi.org/10.1103/RevModPhys.89.015003} {\bibfield  {journal}
  {\bibinfo  {journal} {Reviews of Modern Physics}\ }\textbf {\bibinfo {volume}
  {89}},\ \bibinfo {pages} {015003} (\bibinfo {year} {2017})}\BibitemShut
  {NoStop}%
\bibitem [{\citenamefont {Pankove}(1975)}]{Pankove1975}%
  \BibitemOpen
  \bibfield  {author} {\bibinfo {author} {\bibfnamefont {J.~I.}\ \bibnamefont
  {Pankove}},\ }\href@noop {} {\emph {\bibinfo {title} {Optical processes in
  semiconductors}}}\ (\bibinfo  {publisher} {Courier Corporation},\ \bibinfo
  {year} {1975})\BibitemShut {NoStop}%
\bibitem [{\citenamefont {Rajkanan}, \citenamefont {Singh},\ and\ \citenamefont
  {Shewchun}(1979)}]{RAJKANAN1979}%
  \BibitemOpen
  \bibfield  {author} {\bibinfo {author} {\bibfnamefont {K.}~\bibnamefont
  {Rajkanan}}, \bibinfo {author} {\bibfnamefont {R.}~\bibnamefont {Singh}},\
  and\ \bibinfo {author} {\bibfnamefont {J.}~\bibnamefont {Shewchun}},\
  }\bibfield  {title} {\enquote {\bibinfo {title} {Absorption coefficient of
  silicon for solar cell calculations},}\ }\href
  {https://doi.org/https://doi.org/10.1016/0038-1101(79)90128-X} {\bibfield
  {journal} {\bibinfo  {journal} {Solid-State Electronics}\ }\textbf {\bibinfo
  {volume} {22}},\ \bibinfo {pages} {793--795} (\bibinfo {year}
  {1979})}\BibitemShut {NoStop}%
\bibitem [{\citenamefont {Madelung}(1991)}]{Madelung1991}%
  \BibitemOpen
  \bibfield  {author} {\bibinfo {author} {\bibfnamefont {O.}~\bibnamefont
  {Madelung}},\ }\href
  {https://doi.org/https://doi.org/10.1007/978-3-642-45681-7} {\emph {\bibinfo
  {title} {Semiconductors: group IV elements and III-V compounds}}}\ (\bibinfo
  {publisher} {Springer, Berlin, Heidelberg},\ \bibinfo {year}
  {1991})\BibitemShut {NoStop}%
\bibitem [{\citenamefont {Gonze}\ \emph {et~al.}(2020)\citenamefont {Gonze},
  \citenamefont {Amadon}, \citenamefont {Antonius}, \citenamefont {Arnardi},
  \citenamefont {Baguet}, \citenamefont {Beuken}, \citenamefont {Bieder},
  \citenamefont {Bottin}, \citenamefont {Bouchet}, \citenamefont {Bousquet}
  \emph {et~al.}}]{Gonze2020}%
  \BibitemOpen
  \bibfield  {author} {\bibinfo {author} {\bibfnamefont {X.}~\bibnamefont
  {Gonze}}, \bibinfo {author} {\bibfnamefont {B.}~\bibnamefont {Amadon}},
  \bibinfo {author} {\bibfnamefont {G.}~\bibnamefont {Antonius}}, \bibinfo
  {author} {\bibfnamefont {F.}~\bibnamefont {Arnardi}}, \bibinfo {author}
  {\bibfnamefont {L.}~\bibnamefont {Baguet}}, \bibinfo {author} {\bibfnamefont
  {J.-M.}\ \bibnamefont {Beuken}}, \bibinfo {author} {\bibfnamefont
  {J.}~\bibnamefont {Bieder}}, \bibinfo {author} {\bibfnamefont
  {F.}~\bibnamefont {Bottin}}, \bibinfo {author} {\bibfnamefont
  {J.}~\bibnamefont {Bouchet}}, \bibinfo {author} {\bibfnamefont
  {E.}~\bibnamefont {Bousquet}}, \emph {et~al.},\ }\bibfield  {title} {\enquote
  {\bibinfo {title} {The abinit project: Impact, environment and recent
  developments},}\ }\href
  {https://doi.org/https://doi.org/10.1016/j.cpc.2019.107042} {\bibfield
  {journal} {\bibinfo  {journal} {Computer Physics Communications}\ }\textbf
  {\bibinfo {volume} {248}},\ \bibinfo {pages} {107042} (\bibinfo {year}
  {2020})}\BibitemShut {NoStop}%
\bibitem [{\citenamefont {Romero}\ \emph {et~al.}(2020)\citenamefont {Romero},
  \citenamefont {Allan}, \citenamefont {Amadon}, \citenamefont {Antonius},
  \citenamefont {Applencourt}, \citenamefont {Baguet}, \citenamefont {Bieder},
  \citenamefont {Bottin}, \citenamefont {Bouchet}, \citenamefont {Bousquet}
  \emph {et~al.}}]{Romero2020}%
  \BibitemOpen
  \bibfield  {author} {\bibinfo {author} {\bibfnamefont {A.~H.}\ \bibnamefont
  {Romero}}, \bibinfo {author} {\bibfnamefont {D.~C.}\ \bibnamefont {Allan}},
  \bibinfo {author} {\bibfnamefont {B.}~\bibnamefont {Amadon}}, \bibinfo
  {author} {\bibfnamefont {G.}~\bibnamefont {Antonius}}, \bibinfo {author}
  {\bibfnamefont {T.}~\bibnamefont {Applencourt}}, \bibinfo {author}
  {\bibfnamefont {L.}~\bibnamefont {Baguet}}, \bibinfo {author} {\bibfnamefont
  {J.}~\bibnamefont {Bieder}}, \bibinfo {author} {\bibfnamefont
  {F.}~\bibnamefont {Bottin}}, \bibinfo {author} {\bibfnamefont
  {J.}~\bibnamefont {Bouchet}}, \bibinfo {author} {\bibfnamefont
  {E.}~\bibnamefont {Bousquet}}, \emph {et~al.},\ }\bibfield  {title} {\enquote
  {\bibinfo {title} {Abinit: Overview and focus on selected capabilities},}\
  }\href {https://doi.org/https://doi.org/10.1063/1.5144261} {\bibfield
  {journal} {\bibinfo  {journal} {The Journal of chemical physics}\ }\textbf
  {\bibinfo {volume} {152}},\ \bibinfo {pages} {124102} (\bibinfo {year}
  {2020})}\BibitemShut {NoStop}%
\bibitem [{\citenamefont {Van~Setten}\ \emph {et~al.}(2018)\citenamefont
  {Van~Setten}, \citenamefont {Giantomassi}, \citenamefont {Bousquet},
  \citenamefont {Verstraete}, \citenamefont {Hamann}, \citenamefont {Gonze},\
  and\ \citenamefont {Rignanese}}]{Vansetten2018}%
  \BibitemOpen
  \bibfield  {author} {\bibinfo {author} {\bibfnamefont {M.}~\bibnamefont
  {Van~Setten}}, \bibinfo {author} {\bibfnamefont {M.}~\bibnamefont
  {Giantomassi}}, \bibinfo {author} {\bibfnamefont {E.}~\bibnamefont
  {Bousquet}}, \bibinfo {author} {\bibfnamefont {M.~J.}\ \bibnamefont
  {Verstraete}}, \bibinfo {author} {\bibfnamefont {D.~R.}\ \bibnamefont
  {Hamann}}, \bibinfo {author} {\bibfnamefont {X.}~\bibnamefont {Gonze}},\ and\
  \bibinfo {author} {\bibfnamefont {G.-M.}\ \bibnamefont {Rignanese}},\
  }\bibfield  {title} {\enquote {\bibinfo {title} {The pseudodojo: Training and
  grading a 85 element optimized norm-conserving pseudopotential table},}\
  }\href {https://doi.org/https://doi.org/10.1016/j.cpc.2018.01.012} {\bibfield
   {journal} {\bibinfo  {journal} {Computer Physics Communications}\ }\textbf
  {\bibinfo {volume} {226}},\ \bibinfo {pages} {39--54} (\bibinfo {year}
  {2018})}\BibitemShut {NoStop}%
\bibitem [{\citenamefont {Monkhorst}\ and\ \citenamefont
  {Pack}(1976)}]{Monkhorst1976}%
  \BibitemOpen
  \bibfield  {author} {\bibinfo {author} {\bibfnamefont {H.~J.}\ \bibnamefont
  {Monkhorst}}\ and\ \bibinfo {author} {\bibfnamefont {J.~D.}\ \bibnamefont
  {Pack}},\ }\bibfield  {title} {\enquote {\bibinfo {title} {Special points for
  brillouin-zone integrations},}\ }\href
  {https://doi.org/https://doi.org/10.1103/PhysRevB.13.5188} {\bibfield
  {journal} {\bibinfo  {journal} {Physical Review B}\ }\textbf {\bibinfo
  {volume} {13}},\ \bibinfo {pages} {5188} (\bibinfo {year}
  {1976})}\BibitemShut {NoStop}%
\bibitem [{\citenamefont {Gonze}\ and\ \citenamefont {Lee}(1997)}]{Gonze1997a}%
  \BibitemOpen
  \bibfield  {author} {\bibinfo {author} {\bibfnamefont {X.}~\bibnamefont
  {Gonze}}\ and\ \bibinfo {author} {\bibfnamefont {C.}~\bibnamefont {Lee}},\
  }\bibfield  {title} {\enquote {\bibinfo {title} {Dynamical matrices, born
  effective charges, dielectric permittivity tensors, and interatomic force
  constants from density-functional perturbation theory},}\ }\href
  {https://doi.org/https://doi.org/10.1103/PhysRevB.55.10355} {\bibfield
  {journal} {\bibinfo  {journal} {Physical Review B}\ }\textbf {\bibinfo
  {volume} {55}},\ \bibinfo {pages} {10355} (\bibinfo {year}
  {1997})}\BibitemShut {NoStop}%
\bibitem [{\citenamefont {Gonze}(1997)}]{Gonze1997b}%
  \BibitemOpen
  \bibfield  {author} {\bibinfo {author} {\bibfnamefont {X.}~\bibnamefont
  {Gonze}},\ }\bibfield  {title} {\enquote {\bibinfo {title} {First-principles
  responses of solids to atomic displacements and homogeneous electric fields:
  Implementation of a conjugate-gradient algorithm},}\ }\href
  {https://doi.org/https://doi.org/10.1103/PhysRevB.55.10337} {\bibfield
  {journal} {\bibinfo  {journal} {Physical Review B}\ }\textbf {\bibinfo
  {volume} {55}},\ \bibinfo {pages} {10337} (\bibinfo {year}
  {1997})}\BibitemShut {NoStop}%
\bibitem [{\citenamefont {Trimarchi}\ \emph {et~al.}(2011)\citenamefont
  {Trimarchi}, \citenamefont {Peng}, \citenamefont {Im}, \citenamefont
  {Freeman}, \citenamefont {Cloet}, \citenamefont {Raw}, \citenamefont
  {Poeppelmeier}, \citenamefont {Biswas}, \citenamefont {Lany},\ and\
  \citenamefont {Zunger}}]{Trimarchi2011}%
  \BibitemOpen
  \bibfield  {author} {\bibinfo {author} {\bibfnamefont {G.}~\bibnamefont
  {Trimarchi}}, \bibinfo {author} {\bibfnamefont {H.}~\bibnamefont {Peng}},
  \bibinfo {author} {\bibfnamefont {J.}~\bibnamefont {Im}}, \bibinfo {author}
  {\bibfnamefont {A.~J.}\ \bibnamefont {Freeman}}, \bibinfo {author}
  {\bibfnamefont {V.}~\bibnamefont {Cloet}}, \bibinfo {author} {\bibfnamefont
  {A.}~\bibnamefont {Raw}}, \bibinfo {author} {\bibfnamefont {K.~R.}\
  \bibnamefont {Poeppelmeier}}, \bibinfo {author} {\bibfnamefont
  {K.}~\bibnamefont {Biswas}}, \bibinfo {author} {\bibfnamefont
  {S.}~\bibnamefont {Lany}},\ and\ \bibinfo {author} {\bibfnamefont
  {A.}~\bibnamefont {Zunger}},\ }\bibfield  {title} {\enquote {\bibinfo {title}
  {Using design principles to systematically plan the synthesis of
  hole-conducting transparent oxides: Cu 3 vo 4 and ag 3 vo 4 as a case
  study},}\ }\href {https://doi.org/https://doi.org/10.1103/PhysRevB.84.165116}
  {\bibfield  {journal} {\bibinfo  {journal} {Physical Review B}\ }\textbf
  {\bibinfo {volume} {84}},\ \bibinfo {pages} {165116} (\bibinfo {year}
  {2011})}\BibitemShut {NoStop}%
\bibitem [{\citenamefont {Grosso}\ and\ \citenamefont
  {Pastori~Parravicini}(2000)}]{Grosso2000}%
  \BibitemOpen
  \bibfield  {author} {\bibinfo {author} {\bibfnamefont {G.}~\bibnamefont
  {Grosso}}\ and\ \bibinfo {author} {\bibfnamefont {G.}~\bibnamefont
  {Pastori~Parravicini}},\ }\href@noop {} {\emph {\bibinfo {title} {Solid State
  Physics}}},\ \bibinfo {edition} {1st}\ ed.\ (\bibinfo  {publisher} {Academic
  Press},\ \bibinfo {address} {New York},\ \bibinfo {year} {2000})\BibitemShut
  {NoStop}%
\bibitem [{\citenamefont {Green}(1990)}]{Green1990}%
  \BibitemOpen
  \bibfield  {author} {\bibinfo {author} {\bibfnamefont {M.~A.}\ \bibnamefont
  {Green}},\ }\bibfield  {title} {\enquote {\bibinfo {title} {Intrinsic
  concentration, effective densities of states, and effective mass in
  silicon},}\ }\href {https://doi.org/https://doi.org/10.1063/1.345414}
  {\bibfield  {journal} {\bibinfo  {journal} {Journal of Applied Physics}\
  }\textbf {\bibinfo {volume} {67}},\ \bibinfo {pages} {2944--2954} (\bibinfo
  {year} {1990})}\BibitemShut {NoStop}%
\bibitem [{\citenamefont {Ridley}(1990)}]{Ridley1990}%
  \BibitemOpen
  \bibfield  {author} {\bibinfo {author} {\bibfnamefont {B.~K.}\ \bibnamefont
  {Ridley}},\ }\href@noop {} {\emph {\bibinfo {title} {Quantum processes in
  semiconductors}}},\ \bibinfo {edition} {4th}\ ed.\ (\bibinfo  {publisher}
  {Oxford University Press},\ \bibinfo {year} {1990})\BibitemShut {NoStop}%
\bibitem [{\citenamefont {Vandenberghe}\ and\ \citenamefont
  {Fischetti}(2015)}]{Vandenberghe2015}%
  \BibitemOpen
  \bibfield  {author} {\bibinfo {author} {\bibfnamefont {W.~G.}\ \bibnamefont
  {Vandenberghe}}\ and\ \bibinfo {author} {\bibfnamefont {M.~V.}\ \bibnamefont
  {Fischetti}},\ }\bibfield  {title} {\enquote {\bibinfo {title} {Deformation
  potentials for band-to-band tunneling in silicon and germanium from first
  principles},}\ }\href {https://doi.org/https://doi.org/10.1063/1.4905591}
  {\bibfield  {journal} {\bibinfo  {journal} {Applied Physics Letters}\
  }\textbf {\bibinfo {volume} {106}},\ \bibinfo {pages} {013505} (\bibinfo
  {year} {2015})}\BibitemShut {NoStop}%
\bibitem [{\citenamefont {Barber}(1967)}]{Barber1967}%
  \BibitemOpen
  \bibfield  {author} {\bibinfo {author} {\bibfnamefont {H.}~\bibnamefont
  {Barber}},\ }\bibfield  {title} {\enquote {\bibinfo {title} {Effective mass
  and intrinsic concentration in silicon},}\ }\href
  {https://doi.org/https://doi.org/10.1016/0038-1101(67)90122-0} {\bibfield
  {journal} {\bibinfo  {journal} {Solid-State Electronics}\ }\textbf {\bibinfo
  {volume} {10}},\ \bibinfo {pages} {1039--1051} (\bibinfo {year}
  {1967})}\BibitemShut {NoStop}%
\bibitem [{\citenamefont {Ramos}\ \emph {et~al.}(2001)\citenamefont {Ramos},
  \citenamefont {Teles}, \citenamefont {Scolfaro}, \citenamefont {Castineira},
  \citenamefont {Rosa},\ and\ \citenamefont {Leite}}]{Ramos2001}%
  \BibitemOpen
  \bibfield  {author} {\bibinfo {author} {\bibfnamefont {L.~E.}\ \bibnamefont
  {Ramos}}, \bibinfo {author} {\bibfnamefont {L.~K.}\ \bibnamefont {Teles}},
  \bibinfo {author} {\bibfnamefont {L.~M.}\ \bibnamefont {Scolfaro}}, \bibinfo
  {author} {\bibfnamefont {J.~L.}\ \bibnamefont {Castineira}}, \bibinfo
  {author} {\bibfnamefont {A.}~\bibnamefont {Rosa}},\ and\ \bibinfo {author}
  {\bibfnamefont {J.~R.}\ \bibnamefont {Leite}},\ }\bibfield  {title} {\enquote
  {\bibinfo {title} {Structural, electronic, and effective-mass properties of
  silicon and zinc-blende group-iii nitride semiconductor compounds},}\ }\href
  {https://doi.org/https://doi.org/10.1103/PhysRevB.63.165210} {\bibfield
  {journal} {\bibinfo  {journal} {Physical Review B}\ }\textbf {\bibinfo
  {volume} {63}},\ \bibinfo {pages} {165210} (\bibinfo {year}
  {2001})}\BibitemShut {NoStop}%
\bibitem [{\citenamefont {Bouhassoune}\ and\ \citenamefont
  {Schindlmayr}(2010)}]{Bouhassoune2010}%
  \BibitemOpen
  \bibfield  {author} {\bibinfo {author} {\bibfnamefont {M.}~\bibnamefont
  {Bouhassoune}}\ and\ \bibinfo {author} {\bibfnamefont {A.}~\bibnamefont
  {Schindlmayr}},\ }\bibfield  {title} {\enquote {\bibinfo {title} {Electronic
  structure and effective masses in strained silicon},}\ }\href
  {https://doi.org/https://doi.org/10.1002/pssc.200982470} {\bibfield
  {journal} {\bibinfo  {journal} {Physica Status Solidi (c)}\ }\textbf
  {\bibinfo {volume} {7}},\ \bibinfo {pages} {460--463} (\bibinfo {year}
  {2010})}\BibitemShut {NoStop}%
\bibitem [{\citenamefont {Green}\ and\ \citenamefont
  {Keevers}(1995)}]{Green1995}%
  \BibitemOpen
  \bibfield  {author} {\bibinfo {author} {\bibfnamefont {M.~A.}\ \bibnamefont
  {Green}}\ and\ \bibinfo {author} {\bibfnamefont {M.~J.}\ \bibnamefont
  {Keevers}},\ }\bibfield  {title} {\enquote {\bibinfo {title} {Optical
  properties of intrinsic silicon at 300 k},}\ }\href
  {https://doi.org/https://doi.org/10.1002/pip.4670030303} {\bibfield
  {journal} {\bibinfo  {journal} {Progress in Photovoltaics: Research and
  Applications}\ }\textbf {\bibinfo {volume} {3}},\ \bibinfo {pages} {189--192}
  (\bibinfo {year} {1995})}\BibitemShut {NoStop}%
\bibitem [{\citenamefont {Kane}(1966)}]{Kane1966}%
  \BibitemOpen
  \bibfield  {author} {\bibinfo {author} {\bibfnamefont {E.}~\bibnamefont
  {Kane}},\ }\bibfield  {title} {\enquote {\bibinfo {title} {Band structure of
  silicon from an adjusted heine-abarenkov calculation},}\ }\href
  {https://doi.org/https://doi.org/10.1103/PhysRev.146.558} {\bibfield
  {journal} {\bibinfo  {journal} {Physical Review}\ }\textbf {\bibinfo {volume}
  {146}},\ \bibinfo {pages} {558} (\bibinfo {year} {1966})}\BibitemShut
  {NoStop}%
\bibitem [{\citenamefont {Hybertsen}\ and\ \citenamefont
  {Louie}(1986)}]{Hybertsen1986}%
  \BibitemOpen
  \bibfield  {author} {\bibinfo {author} {\bibfnamefont {M.~S.}\ \bibnamefont
  {Hybertsen}}\ and\ \bibinfo {author} {\bibfnamefont {S.~G.}\ \bibnamefont
  {Louie}},\ }\bibfield  {title} {\enquote {\bibinfo {title} {Electron
  correlation in semiconductors and insulators: Band gaps and quasiparticle
  energies},}\ }\href
  {https://doi.org/https://doi.org/10.1103/PhysRevB.34.5390} {\bibfield
  {journal} {\bibinfo  {journal} {Physical Review B}\ }\textbf {\bibinfo
  {volume} {34}},\ \bibinfo {pages} {5390} (\bibinfo {year}
  {1986})}\BibitemShut {NoStop}%
\bibitem [{\citenamefont {Bouhassoune}\ and\ \citenamefont
  {Schindlmayr}(2015)}]{Bouhassoune2015}%
  \BibitemOpen
  \bibfield  {author} {\bibinfo {author} {\bibfnamefont {M.}~\bibnamefont
  {Bouhassoune}}\ and\ \bibinfo {author} {\bibfnamefont {A.}~\bibnamefont
  {Schindlmayr}},\ }\bibfield  {title} {\enquote {\bibinfo {title} {Ab initio
  study of strain effects on the quasiparticle bands and effective masses in
  silicon},}\ }\href {https://doi.org/https://doi.org/10.1155/2015/453125}
  {\bibfield  {journal} {\bibinfo  {journal} {Advances in condensed matter
  physics}\ }\textbf {\bibinfo {volume} {2015}},\ \bibinfo {pages} {453125}
  (\bibinfo {year} {2015})}\BibitemShut {NoStop}%
\end{thebibliography}%

\end{document}